\RequirePackage{fix-cm}
\documentclass[smallextended]{svjour3}       
\smartqed  
\usepackage{graphicx}
\usepackage{amssymb,amsfonts,amsmath}
\usepackage{url}
\newcommand{\df}{\bf}
\newcommand{\pure}{{\rm pure}}
\newcommand{\applied}{{\rm applied}}
\begin{document}

\title{Evolutionary Events in a Mathematical Sciences Research Collaboration Network\thanks{J. C. Brunson, S. Fassino, A. McInnes, M. Narayan, and B. Richardson were partially funded by NSF Award:477855.
A. McInnes and B. Richardson were partially funded by HHMI:52006309.
S. Fassino, A. McInnes, M. Narayan, and B. Richardson contributed equally.}
}


\author{Jason Cory Brunson \and Steve Fassino \and Antonio McInnes \and Monisha Narayan \and Brianna Richardson \and Christopher Franck \and Patrick Ion \and Reinhard Laubenbacher
}


\institute{J. C. Brunson \at
              Virginia Bioinformatics Institute,
              Washington St, MC 0477, Virginia Tech, Blacksburg, VA  24061 \\
           \and
           S. Fassino \at
              Department of Mathematics,
              227 Ayres Hall, University of Tennessee, Knoxville, TN  37996
           \and
           A. McInnes \at
              Department of Mathematics and Computer Science,
              Oakwood University, Cooper Complex Bld. B, 7000 Adventist Blvd, Huntsville, AL  35896
           \and
           M. Narayan \at
              Lyman Briggs College,
              Michigan State University, 35 East Holmes Hall, East Lansing, MI  48825
           \and
           B. Richardson \at
              Department of Mathematics and Computer Science,
              Oakwood University, Cooper Complex Bld. B, 7000 Adventist Blvd, Huntsville, AL  35896
           \and
           C. Franck \at
              Laboratory for Interdisciplinary Statistical Analysis,
              212 Hutcheson Hall, Blacksburg, VA  24061
           \and
           P. Ion \at
              Mathematical Reviews,
              P.O. Box 8604, Ann Arbor, MI  48107
           \and
           R. Laubenbacher \at
              Center for Quantitative Medicine,
              University of Connecticut Health Center,
              195 Farmington Ave,
              Farmington, CT  06030
              Tel.: +186-04-367516\\
              \email{laubenbacher@uchc.edu}           
}

\date{Received: date / Accepted: date}

\maketitle

\begin{abstract}
This study examines long-term trends and shifting behavior in the collaboration network of mathematics literature, using a subset of data from {\it Mathematical Reviews} spanning 1985--2009.
Rather than modeling the network cumulatively, this study traces the evolution of the ``here and now'' using fixed-duration sliding windows.
The analysis uses a suite of common network diagnostics, including the distributions of degrees, distances, and clustering, to track network structure.
Several random models that call these diagnostics as parameters help tease them apart as factors from the values of others.
Some behaviors are consistent over the entire interval, but most diagnostics indicate that the network's structural evolution is dominated by occasional dramatic shifts in otherwise steady trends.
These behaviors are not distributed evenly across the network; stark differences in evolution can be observed between two major subnetworks, loosely thought of as ``pure'' and ``applied'', which approximately partition the aggregate.
The paper characterizes two major events along the mathematics network trajectory and discusses possible explanatory factors.
\end{abstract}

\section{Introduction}\label{intro}

The evolution of real-world networks, particularly social networks, has been of rising interest.
As time-resolved databases of scientific literature (and of other network-theoretic data) have grown in size and duration, increasingly perceptive diagnostics and rich models of network behavior have been developed \cite{gh-matrix,hs-temporal}.
Most of these studies have investigated limiting behavior in network structure, such as average distance and clustering, or consistencies in network evolution, such as preferential attachment and transitive closure \cite{bjnrsv-evolution,n-structure,n-coauthorship,tl-empirical}.
In contrast, in this paper we investigate the irregularities in the evolution of a collaboration network.

We draw our data, spanning a quarter-century, from the {\tt MathSciNet} database, which consists of publication records from the secondary journal {\it Mathematical Reviews} (MR) published by the American Mathematical Society.
We study the evolution of the network with respect to several well-established diagnostics and distributions, both in raw form, for meaningful comparison to other collaboration networks, and relative to the predictions of several popular random graph models.
While mathematics is as methodologically mature a discipline as any, it is widely viewed as a solitary, or minimally collaborative, enterprise.
Mathematics collaboration networks have been shown to exhibit lower connectivity than other scholarship networks \cite{n-scientific-I}, but, as in other disciplines, there has been discussion of rising collaborativeness in mathematics \cite{g-evolution}, the characterization of which may be viewed as a central goal of this study.

Evolving collaboration networks have been modeled graph-theoretically in three principal ways (our terminology): the {\it cumulative} model that compiles a network incrementally over time from a fixed beginning \cite{bjnrsv-evolution,tl-empirical}, the {\it active} model consisting of a sequence of graphs constructed across several comparable intervals of time \cite{glm-economics,g-evolution}, and the {\it temporal} model that represents the collaboration network as a single time-resolved structure \cite{hs-temporal}.
We require for our analysis a model that can be viewed locally in time, which precludes a cumulative model; this is just as well, since our data by no means trace to the inception of mathematics publishing.
Whereas we are not interested in the careers of individual mathematicians, we do not require the comprehensive (and memory-intensive) temporal model.

The paper is organized as follows:
Section~\ref{design} describes the data we use and the graph-theoretic approach we take.
Section~\ref{results} consists of several subsections in which we analyze specific structural properties such as connectivity, distance, and clustering.
We interpret these analyses, consider possible real-world factors, and suggest further avenues of research in Section~\ref{discussion}, and we wrap up the exposition in Section~\ref{conclusion}.

\section{Design}\label{design}

\subsection{Motivating questions}\label{questions}

Our study addresses three overarching questions:
\begin{enumerate}
\item How does the network evolve, and what irregularities punctuate this evolution?
\item How does the collaboration network of authors in the mathematical sciences compare to other collaboration networks?
\item How do collaborative trends differ across subdisciplines within the mathematical sciences?
\end{enumerate}
In each subsection of Section~\ref{results} we describe the structural properties we intend to trace over time, then present and discuss the results in the context of these questions.
At each step we build upon the previous steps, for instance by invoking maximum-entropy models of the network determined by previously evaluated diagnostics (such as the Erd\H{o}s--R\'{e}nyi model after evaluating the network size and density), or by analyzing the time series themselves (change point analysis, last section).

\subsection{Data}\label{data}

The MR database contains bibliographic information on publications tracing back to 1940.
We extracted, for each entry published within the time period 1985--2009,
an encoded publication index, the year of publication, an encoded ID for each author (consistent throughout the database), and the subject classification(s) assigned to the publication by MR editors.\footnote{These classifications are increasingly often suggested by authors and reviewers but are ultimately decided upon by the editors.}
Our extracted data includes nearly 1.6 million publications that credit nearly 430,000 authors.
We study these data as a proxy for the mathematics literature over this time period.

Things carries many caveats.
MR takes pains to address many of these, and is probably as complete and correct as any scientific publication database subject to its reach.
For instance, MR solicits mathematics literature across countries and languages \cite{j-Chinese} and takes steps to reconcile different naming conventions for common authors \cite{gi-portion}.
However, not only what mathematics literature is excluded from MR but what other literature is written by authors who appear in this database will be absent from this analysis.
See Ref.~\cite{g-coauthorship} for a thorough discussion of such considerations.
Additionally, a recent analysis of the Science Citation Index (SCI) reveals that the database accounts for a decreasing proportion of the total scientific output.
The same trend could be at work here, rendering MR a gradually less complete proxy for the mathematics literature.
Such possibilities are not our focus, but we will remain conscious of them.

\begin{table}[t]
\caption{The MR network over two intervals.}
\label{MRs}
\centering
\begin{tabular*}{\columnwidth}{lrr}
\hline\noalign{\smallskip}
{\it MR} network & 1940--2000 \cite{g-evolution,n-coauthorship} & 1985--2009 \\
\noalign{\smallskip}\hline\noalign{\smallskip}
years & 61 & 25 \\
papers \(p\) & 1598 & 1599 \\
authors \(n\) & 337 & 429 \\
avg. authors/paper \(\overline{a}\) & 1.45 & 1.75 \\ 
avg. papers/author \(\overline{q}\) & 6.9 & 6.5 \\
collab. pairs \(m\) & 496 & 876 \\
avg. no. coauthors \(\overline{k}\) & 2.9 & 4.1 \\
prop. in largest comp. \(n_1/n\)& .62 & .75 \\
avg. separation \(\overline{d}\) & 7.56 & 7.31 \\
global clustering coeff. \(C\) & .15 & .14 \\
avg. clustering coeff. \(\overline{c}\) & .34 & .61 \\
\noalign{\smallskip}\hline
\end{tabular*}
\end{table}

Other subsets of data extracted from the MR database have been studied graph-theoretically \cite{cl-average,csn-power-law,g-evolution,gi-portion,li-rate,p-little,sv-network}.
Table~\ref{MRs} compares several calculations performed on the cumulative network from 1940 to 2000 \cite{g-evolution} and their equivalents on that from 1985 to 2009 (present paper).
These will be discussed in more detail in the next section.
The comparisons are not strictly appropriate due to the different durations over which networks are constructed, but the comparison heralds trends we observe within our 25-year interval---some that have been observed in many collaboration networks, such as toward more, and more frequent, coauthorship, and others that have not, to our knowledge, been described elsewhere, for instance an increasing proportion of authors in the largest component.

\subsection{Models and methods}\label{methods}

We modeled the MR network as a graph in two ways.
The two-mode {\df attribution graph} \(G_2=(P,N,E_2)\) consists of nodes of two ``modes'': the set \(N\) corresponding to researchers and the set \(P\) corresponding to publications.
Each edge \((i,j)\in E_2\subset N\times P\) indicates that publication \(i\) is attributed to researcher \(j\) (among possible others).
The {\df coauthorship graph} \(G_1=(N,E_1)\) is the one-mode projection of \(G_2\) onto \(N\).
It has node set \(N\), and each edge \((j,j')\in E_1\subset N\times N\) indicates that researchers \(i\) and \(j\) have coauthored at least one publication.
A study of the MR attribution graph was an open question from \cite{g-evolution}, in which only the coauthorship graph was scrutinized.

In the next section we present our analysis, organized in sections according to the structural properties being investigated (connectivity, decomposition into components, etc.).
Much of our analysis consists of time series of single-value diagnostics, such as the vertex and edge counts of graphs and their average degrees.
To construct a time series for diagnostic \(D\) over an interval \([a,b]\), take a graph \(G\) and a fixed duration \(\Delta t\).
For each \(t=a+\Delta t,a+\Delta t+1,\ldots,b-1,b\), take \(G_t\) to be the graph constructed over the interval \([t-\Delta t,t]\) and compute \(D(G_t)\).
The time series is then \((D(G_{a+\Delta t}),\ldots,D(G_b))\).
Following the time resolution of the database, we let \(t\) take integer values between \(1984+\Delta t\) and \(2009\), where the value \(t\) corresponds to the moment of changeover from calendar year \(t\) to calendar year \(t+1\).
For example, when \(\Delta t=5\) we get time series of length 21 computed over the intervals 1985--9 through 2005--9.

In addition to the ``aggregate'' network constructed from all publications, we study networks constructed from two subsets of the literature that very nearly divide it in half.
These we determine by splitting the subject classifications into one range that covers mathematics subdisciplines popularly considered more ``pure'' and another more ``applied''.
These classifications are taken from the AMS Mathematics Subject Classification (MSC) scheme, and the ranges are defined at the 2-character prefix level by 03--58 and 60--94, respectively.\footnote{See the MSC itself at \url{http://www.ams.org/mathscinet/msc/msc2010.html} for finer detail.}
The resulting subnetworks receive much the same treatment as the aggregate.
We expect differences in behavior between the pure and applied subnetworks to yield insights into the range and mechanisms of attribution and coauthorship graph structures.

\section{Results}\label{results}


\subsection{Rates of growth, publication, and collaboration}\label{growth}

\begin{figure}[h]
\centerline{\includegraphics[width=\columnwidth]{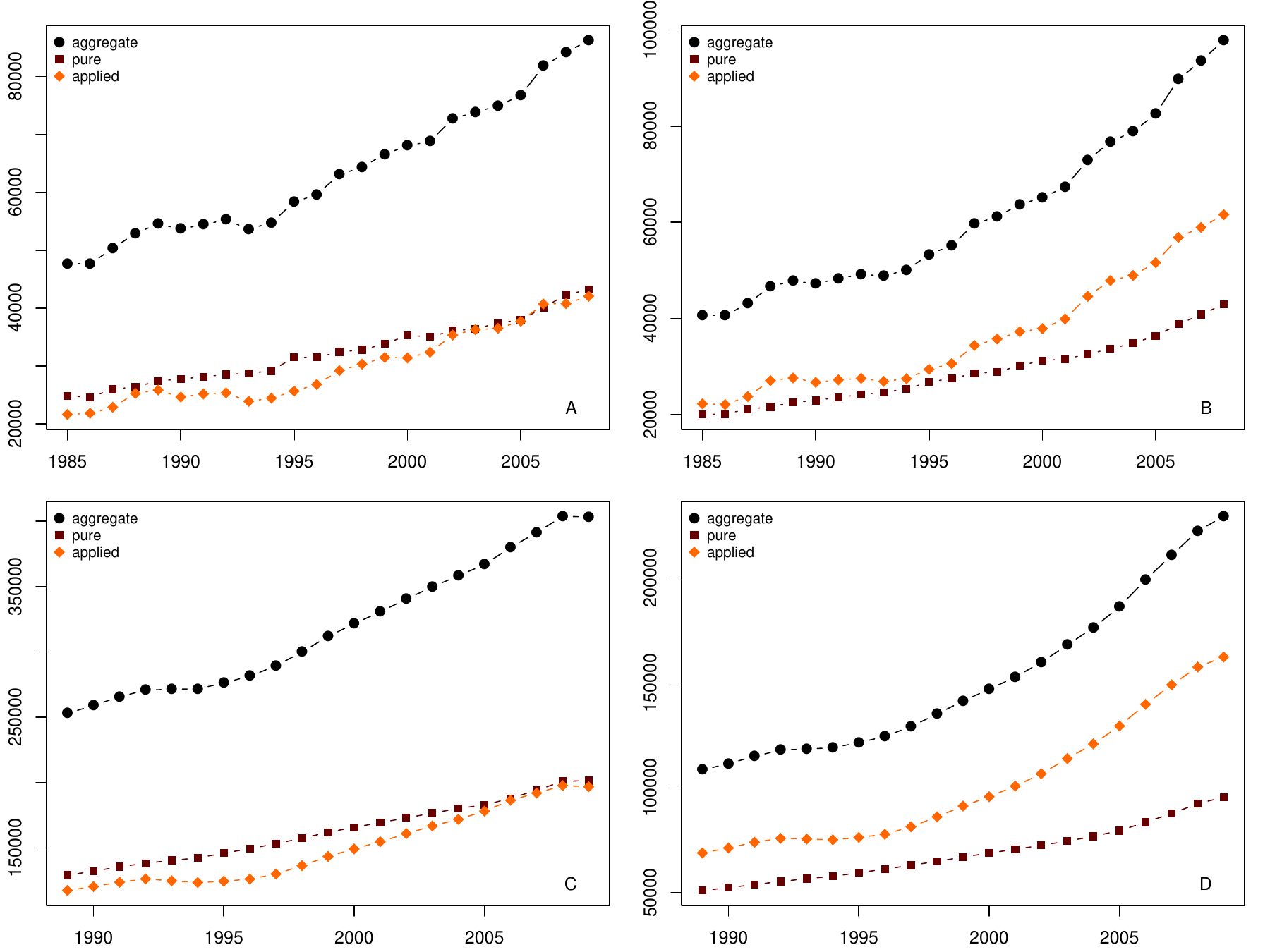}}
\caption{\small For the aggregate, pure, and applied networks at each year,
{\sc(a)} the total number of recorded publications and
{\sc(b)} the total number of attributed authors.
{\sc(c,d)} The same calculations across a 5-year sliding window.\label{fig:size}}
\end{figure}

The active literature compiled by MR and community of authors who produced it have both grown over our 25-year interval, though not monotonically.
The time series for \(p\) and \(n\) are depicted in Fig.~\ref{fig:size}.\footnote{Our data from 2009 is incomplete and so is omitted from the 1-year plots. We include it in 5-year plots and analyses with the expectation that the impact of the missing data on the 5-year calculations will be slight.}
The growth of scientific literatures and communities has traditionally been modeled exponentially \cite{li-rate,pgd-inflationary,p-little}.
For the exponential model
\[x=x_0e^{rt}+\epsilon\]
(with Gaussian errors), we obtained estimated growth factors \(r=.026\) (publications) and \(r=.040\) (researchers), though these models do not fit the data well.\footnote{The numbers are accelerating more rapidly than an exponential growth model can account for, given that the model assumes that \(\lim_{t\to-\infty}x=0\).}

It is notable that, though growth of the literature outpaced that of the community over our interval, the rates of growth of the literature and of the community were very similar over the 60-year interval studied in \cite{g-evolution}:
Fitting the same model to the sizes of the literature and of the community across adjacent decades obtains the very similar growth rates of \(r=.0425\) and \(r=.0433\), while fitting to the data over 10-year windows through our interval obtains \(r=.026\) and \(r=.043\).
For the remainder of the analysis we view these growths as independent parameters.\footnote{Because authors, unlike publications, recur over time, comparisons like these become problematic between intervals of different duration.}

The increasing ratio of researchers to publications, especially after 2000, suggests that collaboration or publication habits---or both---in mathematics have been in flux.
The trend could be explained by a rise in the typical number of authors per publication or by a decline in the typical number of publications per researcher.
These are the degrees of the publication and researcher nodes of \(G_2\), respectively.
We refer to the degree of a publication node \(i\in G_2\) (the number of researchers who authored it) as its {\df cooperativity \(a_i\)}, and the degree of a researcher node \(j\) as its {\df productivity \(q_j\)} \cite{g-coauthorship}.
Their averages \(\overline{a}\) and \(\overline{q}\) are related to \(p\) and \(n\) by
\[p\overline{a}=n\overline{q}\text,\]
where both quantities are equal to the total number of attributions \(b=|E(G_2)|\).
Two other network distributions are often used to quantify collaboration and output:
The degree \(k_j\) of a researcher \(j\in G_1\) is the number of collaborators of \(j\) and reflects \(j\)'s tendency to collaborate; and the number \(w_{jj'}\) of publications coauthored by a pair \((j,j')\in E(G_1)\) of collaborators reflects their contributions.
We call \(k_j\) the {\df connectivity} of \(j\) \cite{g-coauthorship} and \(w_{jj'}\) the {\df collaboration weight} of \(j\) and \(j'\) \cite{n-scientific-ii,n-who}.

Other analyses of professional literature reveal typical distributions of these statistics \cite{bjnrsv-evolution,gs-analyzing,glm-economics,m-structure,n-scientific-I,tl-empirical}.
The average cooperativity ranges from just above 1 (the theoretical minimum) to nearly 10 but typically falls below 5.
Analyses of networks over intervals ranging in length from 5 to 10 years tend to yield an average researcher productivity between 3 and 5 and an average connectivity between 1 and 10.
Longitudinal studies have shown increases in each, though increases in typical productivity have been more mild while increases in cooperativity and connectivity have been more drastic.
We can also look back on Grossman's study of the MR data \cite{g-evolution}, in which the author observes average cooperativity rise from 1.10 over the 1940s to 1.63 over the 1990s, average productivity from 3.41 to 4.97, and average connectivity from 0.49 to 2.84.

\begin{figure}[h]
\centerline{\includegraphics[width=\columnwidth]{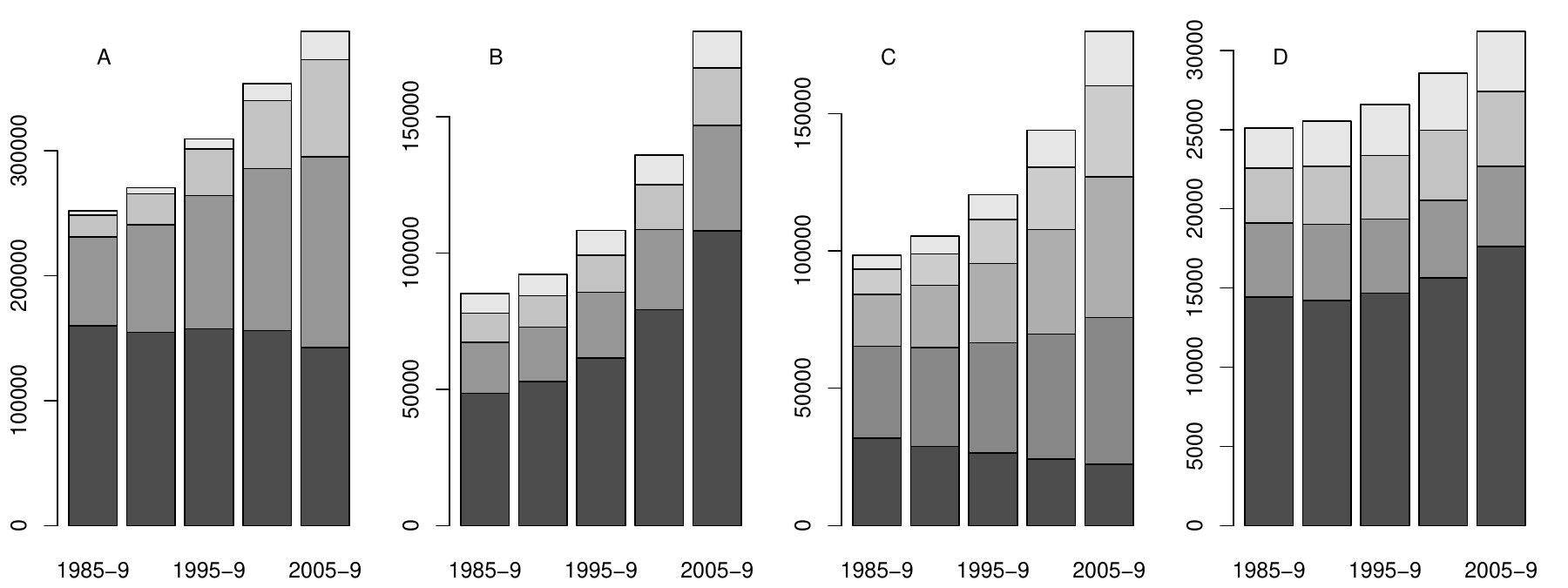}}
\caption{\small Across adjacent 5-year intervals: stratified histograms of {\sc(a)} cooperativity \(a=1,2,3,4\) across publications, {\sc(b)} productivity \(q=1,2,3,4\) across authors, {\sc(c)} connectivity \(k=0,1,2,3\) across authors, and {\sc(d)} collaboration weight \(w=1,2,3,4\) across pairs of coauthors in the aggregate MR network.\label{fig:deg}}
\end{figure}

The stratified histograms of Fig.~\ref{fig:deg} {\sc(a--d)} illustrate the growth and changing composition of the network.
The starkest reallocations occurred within the distributions of cooperativity and connectivity.
The substantial decline of solo (\(k_j=0\)) authors was more than compensated for by the rise in single-collaborator researchers.
The number of solo (\(a_i=0\)) publications remained steady but was greatly diminished in proportion by more cooperative publications.
In both cases the proportional increase was greater for higher values, producing ``fatter-tailed'' distributions.
Mean cooperativity \(\overline{a}\) increased by more than half over the two decades from 1985--9 to 2005--9, while mean connectivity \(\overline{k}\) doubled.
The indicators of publishing frequency---productivity across researchers and collaboration weight across pairs of coauthors---rose only slightly over our interval, and even began to decrease toward the end.
The histograms suggest that this was due to an influx of one-time authors after 2000, which a closer look at the changing proportions of researchers by productivity confirms.

\begin{figure}[h]
\centerline{\includegraphics[width=\columnwidth]{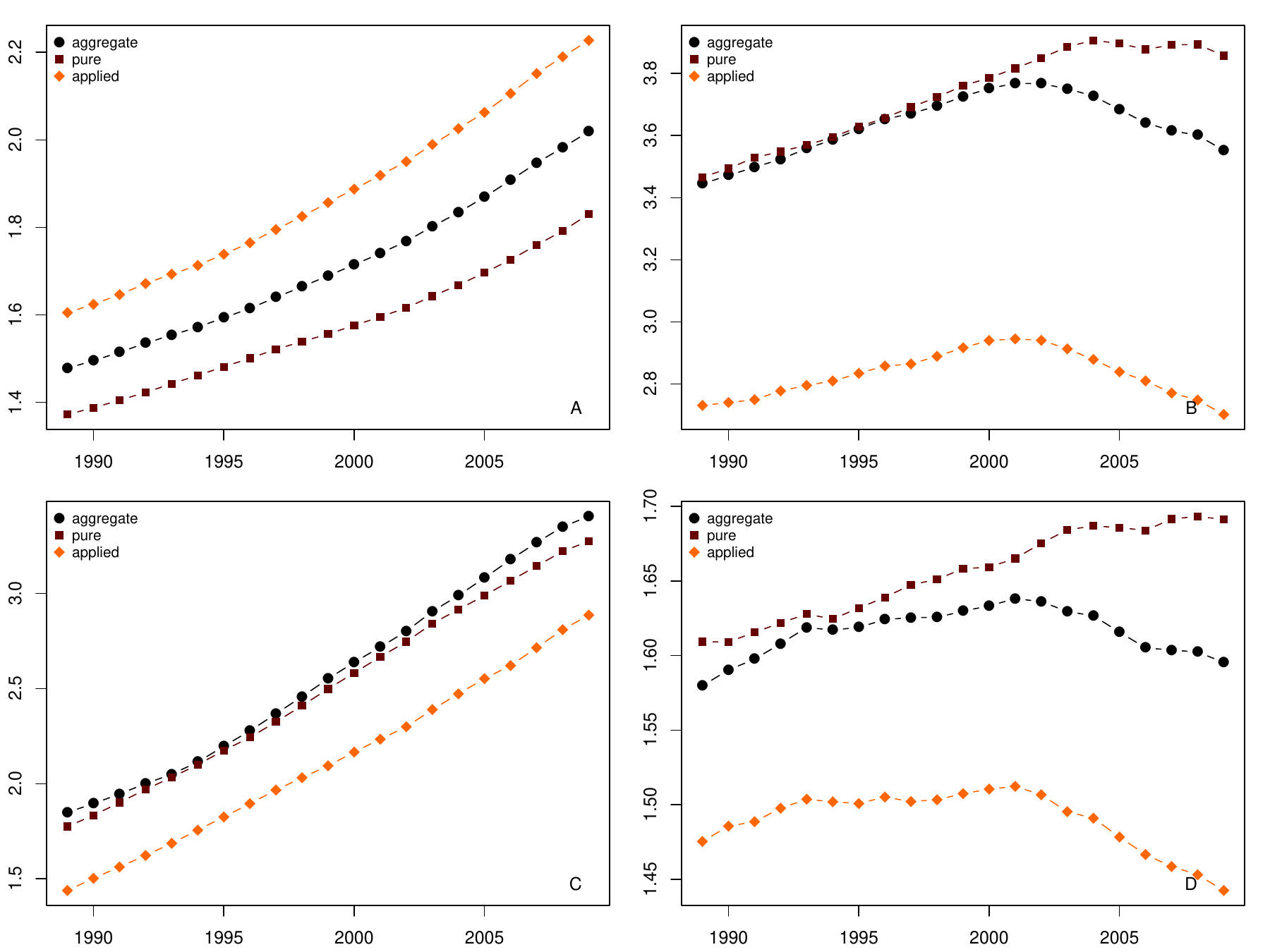}}
\caption{\small Across a sliding 5-year window, arithmetic means of Fig.~\ref{fig:deg} in the aggregate, pure, and applied networks.\label{fig:degavg}}
\end{figure}

The rate of growth of \(\overline{k}\) was approximately piecewise linear; this rate doubled from 1985--94 to 1994--2009, changing pace around the same time that the growth rates of \(P\) and \(N\) noticeably increased.
We refer to the structural phenomenon responsible for this shift as the {\df mid-90s event}.
Later, as acceleration in the numbers \(n\) and \(m\) of researchers and of coauthor pairs accelerated the author-to-publication ratio around 2000, the 5-year averages of \(\overline{q}\) and \(\overline{w}\) abruptly began to decrease.
We refer to this phenomenon as the {\df early-00s event}.
Both shifts were more pronounced in the applied research community, as were the long-term trends:
The applied research community was consistently better-connected in terms of \(a\) and \(k\), however, while the pure was consistently more prolific in terms of \(q\) and \(w\).

\begin{figure}
\centerline{\includegraphics[width=\columnwidth]{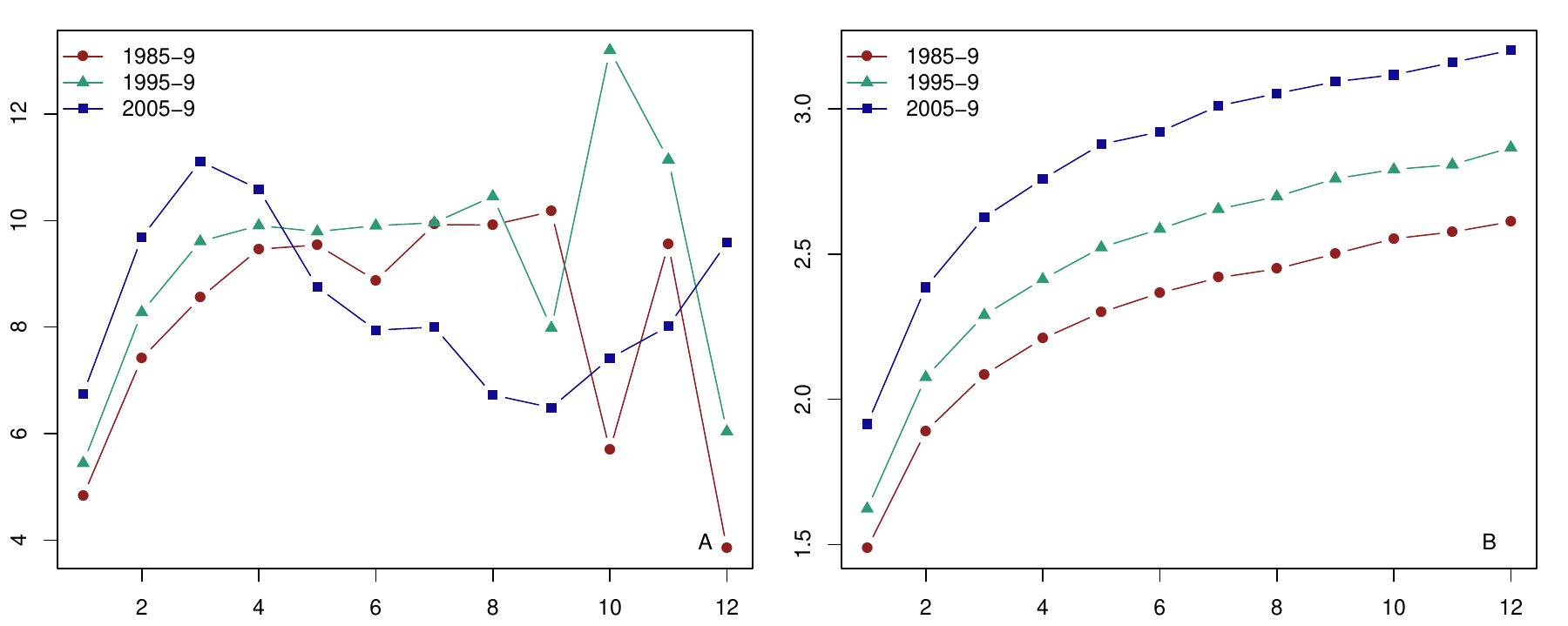}}
\caption{\small For three evenly-spaced 5-year intervals,
{\sc(a)} for values \(a=1,\ldots,12\), the expected mean productivity of the authors on a publication \(i\) with cooperatively \(a_i=a\), and
{\sc(b)} for values \(q=1,\ldots,12\), the expected mean cooperativity of the publications attributed to an author \(j\) with productivity \(q_j=q\).\label{fig:aq}}
\end{figure}

The imbalance of growth between the research community and the published literature is thus due to a more rapid increase in the typical publication's authorship than in the typical author's output.
One natural follow-up consideration is the extent to which prolific researchers tend to be behind the more cooperative publications, or to be more collaborative on average.
The correlation, taken over attributions \((i,j)\in E(G_2)\), between cooperativity and productivity is negligible.\footnote{While always near zero, whether it is positive or negative depends on window size.}
However, the typical cooperativity of a researcher's papers depends positively on that researcher's productivity, {\it and} the typical productivity of a publication's authors depends positively on the publication's cooperativity---to a point.
Fig.~\ref{fig:aq} depicts
\begin{equation}\label{eqn:productivity-vs-cooperativity}
\overline{a}_q\equiv\frac{\sum_{q_j=q}\sum_{(i,j)\in E_B}a_i}{\sum_{q_j=q}q_j}\ \text{versus}\ q
\hspace{.25cm}\text{and}\hspace{.25cm}
\overline{q}_a\equiv\frac{\sum_{a_i=a}\sum_{(i,j)\in E_B}q_j}{\sum_{a_i=a}a_i}\ \text{versus}\ a
\end{equation}
across a 5-year sliding window.\footnote{We may interpret the second expression (\ref{eqn:productivity-vs-cooperativity}) as the expected productivity of a researcher chosen (uniformly) at random from those attributed by a randomly-chosen publication having given cooperativity \(a\), but {\it not} as the expected productivity of a researcher chosen at random from the collection of researchers who have been attributed by {\it some} publication of cooperativity \(a\).}
Both relationships are strongest for small values.
While the former holds for 5-year productivities up to \(q=12\), however, the latter breaks down for cooperativities \(a>4\).
In addition to growing noisier, in more recent years this relationship reversed, so that highly cooperative publications (\(a>4\)) had lower average coauthor productivity than moderately cooperative publications (\(2\leq a\leq 4\)).\footnote{The first plot may be contrasted with Fig.~2 of \cite{g-coauthorship}, which depicts a decline in productivity associated with especially high cooperativity in the mathematics literature (obtained through the SCI), in contrast to the two other scientific literatures in the same study.}

We have uncovered some modest associations among several diagnostics of collaboration and publishing rates, but it is unclear how interdependent these diagnostics are.
Consider the distribution of connectivities \(k_j\) across the nodes of \(G_1\):
How does the distribution differ from what we would expect, knowing only the distributions of cooperativity and productivity in the bipartite \(G_2\)?
How does it differ from the expectations we would form knowing only the size and density of \(G_1\)?
And how much of the structure of \(G_1\) can be attributed to its connectivity distribution?
We adopt three popular random graph models to help answer these questions.

The (uniform) random graph \(G(n,p)\) \cite{er-random}, or ER model after its progenitors, is the distribution arising from assigning an edge between each pair \((j,j')\) of a fixed number \(n\) of nodes with uniform probability \(p\).
The graph has expected density \(p\), while \(G_1\) has density \(\overline{k}/(n-1)\), so to avoid confusion with \(|P|\) we will write \(G(n,\overline{k}/(n-1))\).
This model provides a baseline expectation for \(G_1\) based on size and density alone.
The degree sequence random graph \(G(K)\) \cite{nsw-random}, the NSW model, is distributed uniformly over graphs of a fixed degree sequence \(K=(k_1\geq k_2\geq\cdots)\).
This model arises out of a random rewiring process among nodes that preserves each node's degree.
Since \(K\) determines \(n\) and \(\overline{k}\), the NSW model is strictly narrower than the ER, and provides expectations for other structural properties of \(G_1\) based on the distribution of connectivity.
Finally, an analogous rewiring process that preserves the partition of nodes in a bipartite graph as well as their degrees produces a bipartite NSW (bNSW) model.
This model provides expectations for \(G_2\) but also, via projection, for \(G_1\) based only on the distributions of cooperativity and of productivity.

As an example, we can ask how much of the variation in how widely researchers collaborate is due simply to the sheer number of researchers involved in single projects by comparing the average connectivity of \(G_1\) to its expectation based on the bNSW model.
The latter is given by
\[\overline{k}_{\rm bNSW}=\sum_q\frac{n_q}{n}q\sum_a\frac{p_a}{p}(a-1),\]
where the \(n_q\) and \(p_a\) denote the numbers of researchers and of publications with a given productivity and cooperativity, respectively.
The formula computes the sum of each researcher's connectivity \(q(a-1)\) (under the asymptotic assumption that a researcher's collaborations do not overlap) weighted by its probability \(\frac{n_qp}{np_a}\) (under the underlying assumption that collaborations are independently distributed).

\begin{figure}
\centerline{\includegraphics[width=.5\columnwidth]{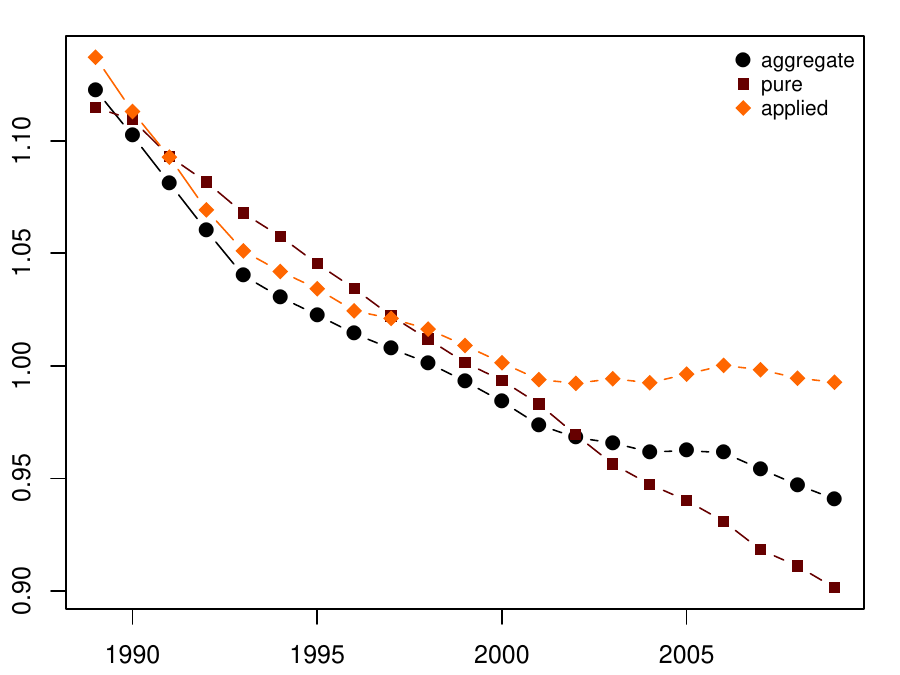}}
\caption{\small Across a 5-year sliding window, the ratio of \(\overline{k}\) to its expectation in the bNSW model.\label{fig:pred}}
\end{figure}

Fig.~\ref{fig:pred} depicts the ratio of \(\overline{k}\) to \(\overline{k}_{\rm bNSW}\) over time.\footnote{We compute the ratio, rather than the difference or another single-value comparison, to better account for the changing size and density of the network.
Optimally, one would compute a test statistic like the \(Z\)-score (e.g.\ \cite{ms-specificity}), but this correctly requires first generating and then running the same (expensive) statistics on a collection of random graphs.}
While the bNSW model provided a consistently close prediction to \(\overline{k}\), this prediction shifted from under- to overestimate over our 25-year interval.
This shift was steady with respect to the pure subnetwork but slowed incrementally with respect to the applied, ceasing after 2000.
The model incorporates cooperativity and productivity so that differences between observations and its predictions reflect cooperativity--productivity correlations and collaborative overlap.
These observations indicate that researchers' families of collaborators shrank, relative to the sheer amount of coauthorship in which researchers engaged, and that this was less true of more applied researchers.
The trend could be due to repeat coauthorship among teams of collaborators or the shifting relationship between cooperativity and productivity, with the pure--applied divide due to an imbalance in either.
We have considered the latter option above and will consider the former in our later discussion of clustering.

\subsection{Multidisciplinarity}\label{multidisciplinarity}

There is a broad recognition that research across or outside established disciplines is becoming more prevalent within the sciences, and the AMS classification scheme offers another lens through which to investigate this trend.
Multidisciplinary, interdisciplinary, and transdisciplinary research trends have been discussed extensively, though the concepts themselves have proven difficult to define \cite{albcfghg-defining,prcp-interdisciplinary}.
In those studies that have compared fields including mathematics, mathematics has tended to be among the less cross-disciplinary \cite{mbg-interdisciplinarity,qla-types}.
Graph-theoretic approaches to quantifying cross-disciplinarity in collaboration networks have been limited \cite{wrbkbkrb-approaches}.

We track cross-disciplinary trends in the MR network in two ways:
First, we use the number \(s_i\geq 1\) of subject classifications assigned to each publication \(i\in P\) as a proxy for the publication's disciplinary breadth.
We adopt for this diagnostic the term ``multidisciplinarity'', the most modest of the above three \cite{albcfghg-defining,wrbkbkrb-approaches}, and we follow the distribution and average of multidisciplinarity over time.
Second, common authorship can be used to establish links among publications in the same way that coauthorship establishes links among researchers:
We define the graph \(G'_1\) in this way.\footnote{We construct \(G'_1\) from the subset of the literature having primary MSC ranging from 03 to 94. The analysis of this unipartite projection of \(G_2\) onto \(P\) rather than \(N\) was another open question from \cite{g-evolution}.}
We ask how much of this connectivity through the literature is between pure and applied publications (as determined by their primary MSC) versus within the pure or applied literatures.
To this end we let \(P_\pure,P_\applied\subset G'_1\) denote the subsets of nodes (publications) having primary MSC in 03--58 and in 60--94, respectively, and define
\[r=\frac{|E(G'_1)|-|E_\pure|-|E_\applied|}{|E(G'_1)|}\text,\]
where \(E_\pure\) and \(E_\applied\) are the subsets of \(E(G'_1)\) that link two pure and two applied publications, respectively.
A baseline is given by
\begin{equation}\label{eqn:pureappliedties}
r_{\rm ER}=\frac{2|P_\pure||P_\applied|}{\textstyle{|P_\pure|+|P_\applied|(|P_\pure|+|P_\applied|-1)}}\text,
\end{equation}
which is the expected value of \(r\) in the absence of preference, given the number of publications of each type.\footnote{We investigated the relationship of multidisciplinarity to cooperativity across publications, analogously to (\ref{eqn:productivity-vs-cooperativity}) though taken over publications rather than attributions, but found no substantive relationship.}

\begin{figure}
\centerline{\includegraphics[width=\columnwidth]{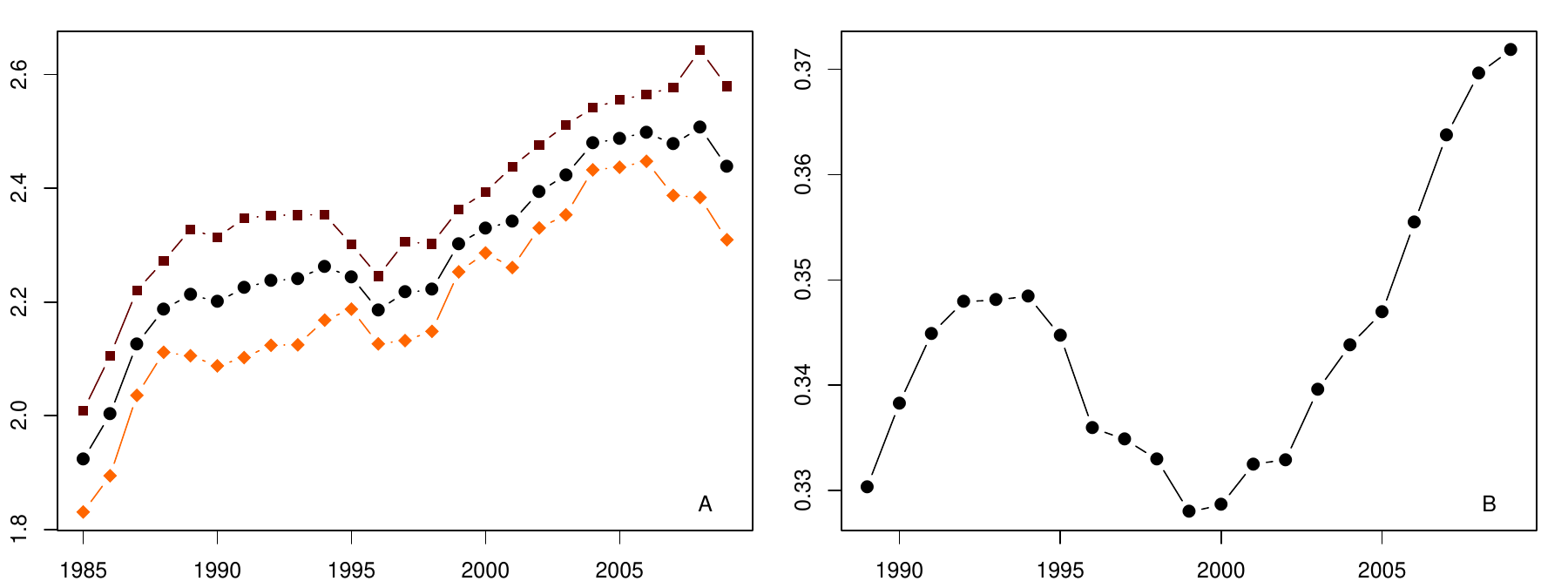}}
\caption{\small {\sc(a)} Across a 5-year sliding window, the average number of subject classifications (including the required one) assigned to a publication.
{\sc(b)} Across years, the proportion of edges in \(G'_1\) that connect pure and applied publications.\label{fig:mult}}
\end{figure}

The MR literature grew increasingly multidisciplinary, and while the pure literature was assigned consistently more classifications on average this trend was shared very closely by both pure and applied literatures.
Meanwhile, the proportion of common authorships that bridge these literatures has been a steady fraction (about a third) of what one would expect based on the rate of common authorships alone.
Both measures of disciplinary interaction weaken over the period 1994--2000 but afterward recover.
This leads us to characterize the mid-90s event by decreased, and the early-00s event by renewed, multidisciplinarity.

\subsection{Connectedness}\label{connectedness}

Absent other factors, as a network grows denser it grows better-connected by other indicators as well.
In this and the following two subsections we'll consider distributions of three such indicators: the sizes of connected components, of internode distances, and of clustering.
We contrast each against the expectations that arise from appropriate random models.
Here we consider the connected components of \(G_1\):
An {\df induced subgraph} \(C\subset G_1\) contains every edge between its nodes that appears in \(G_1\), and \(C\) is a {\df connected component} if it is nonempty, connected (every node can be reached via a path from every other), and maximal as such.

Label the components of \(G_1\) as \(C_1,C_2,\ldots\) in such a way that \(|C_1|\geq|C_2|\geq\cdots\).
As active graphs are constructed over larger durations of time, recording more collaborations among many of the same researchers, an increasing proportion of their nodes will constitute \(C_1\).
Previous research on collaboration networks indicates that this proportion grows into a majority in mature disciplines after three or four years \cite{bjnrsv-evolution,glm-economics,g-evolution,n-scientific-I,p-growth,tl-empirical}.

The ER model exhibits a giant component when \(\sum_jk_j>n\), while the unipartite NSW model has threshold \(\sum_jk^2>2\sum_jk_j\).
In both models \(|C_1|\) scales with \(n\) by a factor that depends on the governing parameters\footnote{This factor derives from \cite{nsw-random} as \(1-\sum_k\frac{n_k}{n}u^k\), where \(u\) is the solution to the equation \(2mu=\sum_kn_kku^{k-1}\) (recall that \(m\) is the number of edges).} while an upper bound on \(|C_2|\) scales similarly with \(\log n\) \cite{er-evolution,mr-critical,mr-size,s-giant}.
\(G_1\) satisfies both thresholds over every 5-year window.

\begin{figure}
\centerline{\includegraphics[width=\columnwidth]{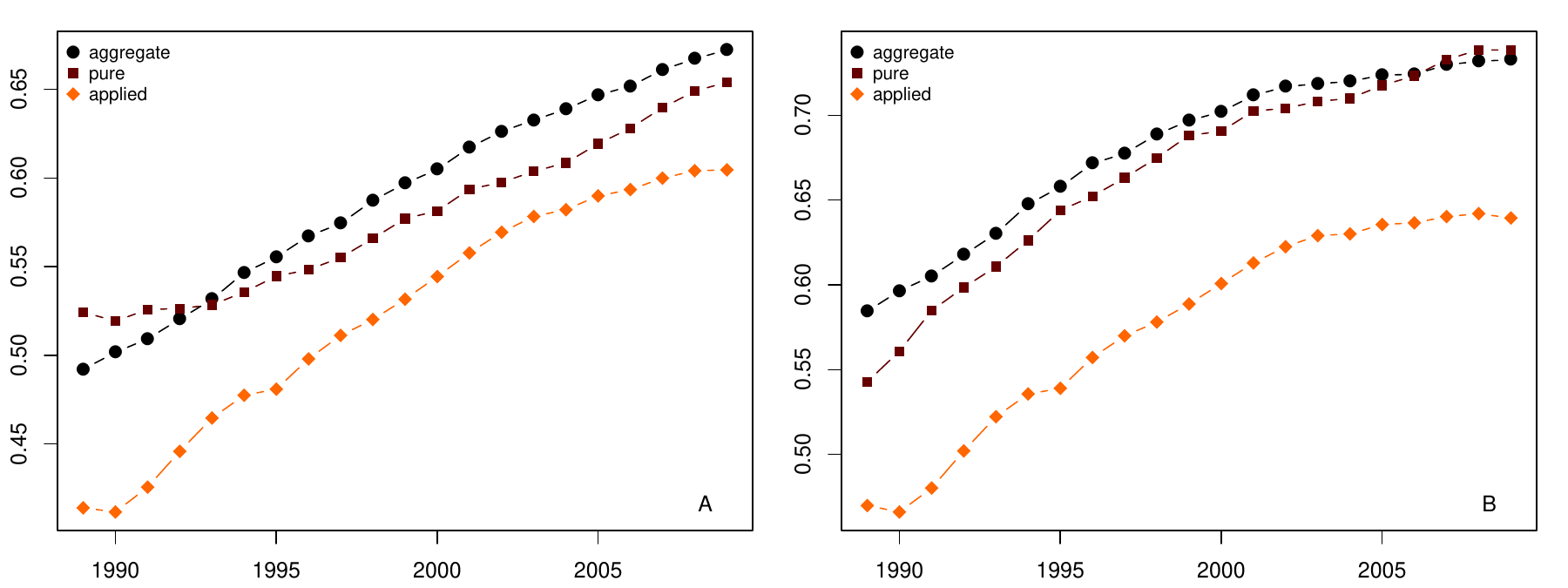}}
\caption{\small The proportion \(|C_1|/n\) of authors in the largest connected component of the coauthorship graph, normalized by its expected values {\sc(a)} in equivalent ER random graphs and {\sc(b)} in equivalent NSW graphs, across a sliding 5-year window.\label{fig:comp}}
\end{figure}

The proportional size of \(C_1\) across 5-year intervals rises from \(37\%\) over 1985--9 to \(65\%\) over 2005--9.
These proportions span the aforecited range of empirical values, which suggests that \(C_1\) has been approaching a practical upper limit.
This observation holds even after size, density, and connectivity are taken into account; \(C_1\) is growing in size {\it in proportion to} the sizes expected from the ER and NSW models, as depicted in Fig.~\ref{fig:comp}.
The connectivity distribution puts constraints on this expectation, in the sense that the expected sizes of \(C_1\) are smaller in the NSW model than in the ER model, but the MR network showed diminishing progress over time in drawing as great a proportion of researchers into a single component as the model achieves through randomness.

We also looked at the distribution of \(|C_i|\) over time (plots not included).
The ratio \(|C_2|/\log n\) maintains a remarkably consistent range of 8 to 10 except over early years of our interval in the applied network.
The size distributions of the non-largest components over each interval very closely follow power laws, as anticipated from previous studies.
The exponent, determined using the power-law fitting method of \cite{csn-power-law} under several fixed starting values of \(k\), likewise shows no consistent trend over time.

\subsection{Distance}\label{distance}

We have seen that the mathematics research community has grown increasingly connected, by a variety of indicators including cooperativity, connectivity/density, and the size distribution of the connected components.
In particular, the increased proportional size of the largest component has outpaced expectations based on the size of the coauthorship graph and its connectivity.
This prompts us to ask whether \(G_1\), and in particular \(C_1\), grew ``better-connected'' by other standards.
Two of the commonest are the typical internode distance and the amount of clustering, the definitional hallmarks of ``small world'' graphs \cite{lm-efficient,ws-collective} and commonly observed features of real-world social, including collaboration, networks \cite{n-scientific-ii}.
In this section we consider the former.
A {\df path} in \(G_1\) is a sequence \((j,j_1,\ldots,j_d)\) of distinct nodes in \(G_1\) each adjacent pair of which form an edge, and the {\df distance} between researchers \(j\) and \(j'\) in \(G_1\) is the minimum length \(d\) of a path from \(j\) to \(j'\).

Network studies typically compute only the average distance \(\overline{d}\) of a network, which calculation omits pairs of nodes that are not connected by a path \cite{bghj-distance,n-structure}.
These averages typically range amidst \(4.6\leq\overline{d}\leq 9.7\) \cite{n-scientific-ii}.
The average distance in an ER graph is known to follow the asymptotic approximation
\[\overline{d}_{\rm ER}\approx\frac{\log n-\gamma}{\log\overline{k}}+\frac{1}{2}\text,\]
where \(\gamma\) is now the Euler--Mascheroni constant \cite{ffh-average}.
Meanwhile, a (unipartite) NSW graph with degree sequence \((k_1\geq\cdots\geq k_n)\) was shown in \cite{cl-average} to have average distance
\[\overline{d}_{\rm NSW}\approx\frac{\log n}{\log(\sum{{k_i}^2}/\sum{k_i})}\text.\]
In both cases the graphs are not necessarily connected.
To assess the average distance in \(G_1\) in light of its density and of its degree sequence, we compute the ratio of \(\overline{d}\) to these expectations for the equivalent ER and NSW graphs over time.

Some studies have taken advantage of the harmonic average distance
\[\overline{d^{-1}}^{-1}=\Big(\sum_{i,j}{d_{ij}}^{-1}/\textstyle{n\choose 2}\Big)^{-1}\]
taken over all pairs of nodes, the reciprocal of the graph's {\df efficiency} \cite{lm-efficient} (see also \cite{oas-node}).
This averaging scheme allocates greater weights to smaller distances.
Additionally, disconnected nodes contribute zero to the sum; the calculation omits no pairs of nodes and thereby detects both distances within components and the disconnectedness of the whole graph.
The relative weights of these is not obvious.
To account for the influence of the components of \(G_1\), we normalize the harmonic average by the value it would take in a graph consisting of components of the same sizes within each of which every internode distance is 1.
This baseline is
\begin{equation}\label{eqn:completecomponents}
\textstyle{n\choose 2}/\sum_c\textstyle{{n_c}\choose 2}=n(n-1)/\sum_c(n_c(n_c-1))\text.
\end{equation}

Finally, we consider the distribution of distances within \(C_1\).
This offers insight into the changing spread of the distribution, unbiased by low distances within smaller components.
The absence of disconnected pairs of researchers in \(C_1\) also permits a meaningful comparison between \(\overline{d}\) and \(\overline{d^{-1}}^{-1}\).

\begin{figure}[h]
\centerline{\includegraphics[width=\columnwidth]{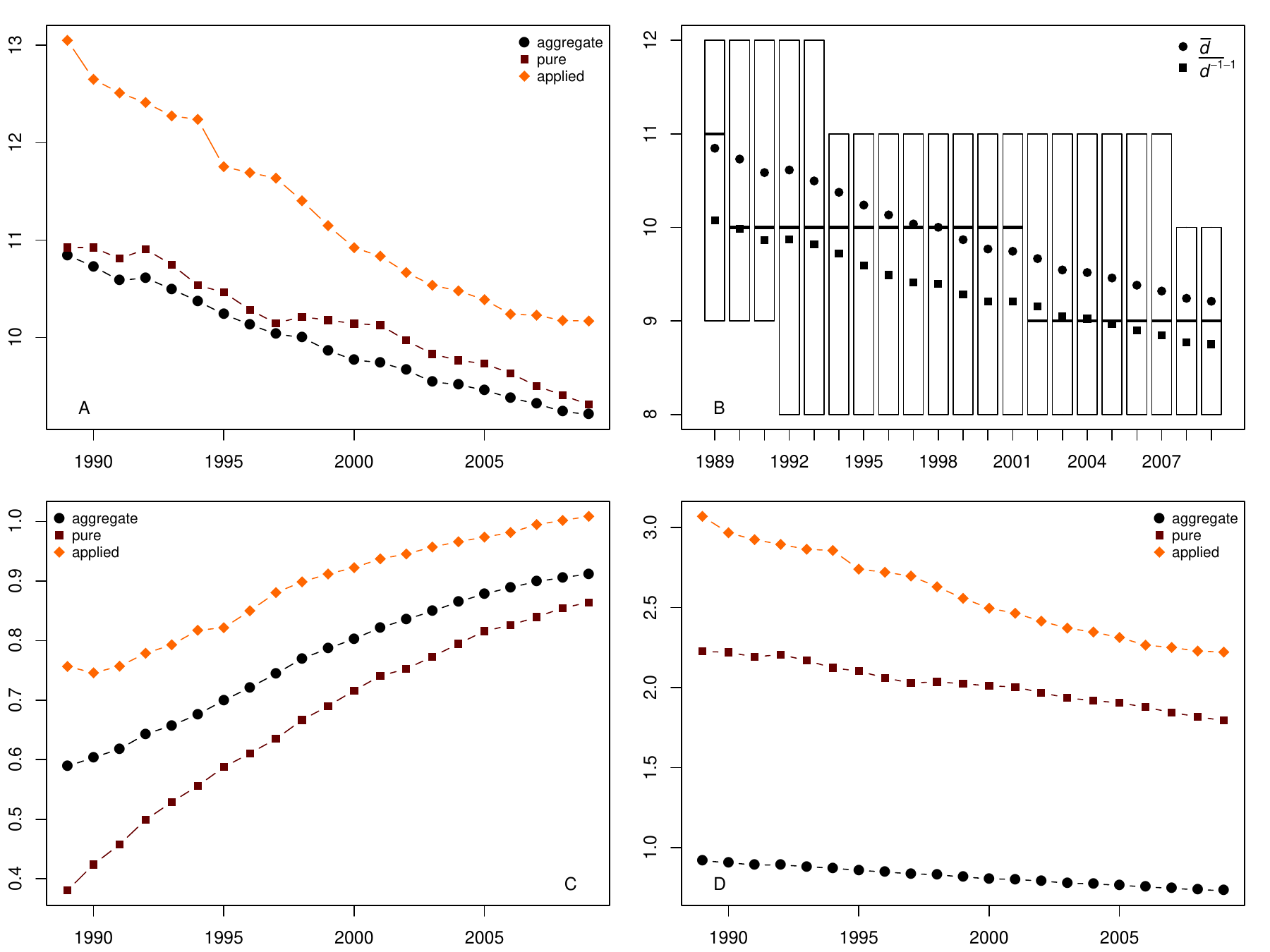}}
\caption{\small {\sc(a)} Mean distance across a 5-year sliding window and {\sc(b)} boxplots of distances within \(C_1\), with arithmetic (circles) and harmonic (squares) means overlaid.
Mean separation normalized {\sc(c)} by ER and {\sc(d)} by NSW predictions.\label{fig:dist}}
\end{figure}

\begin{figure}[h]
\centerline{\includegraphics[width=\columnwidth]{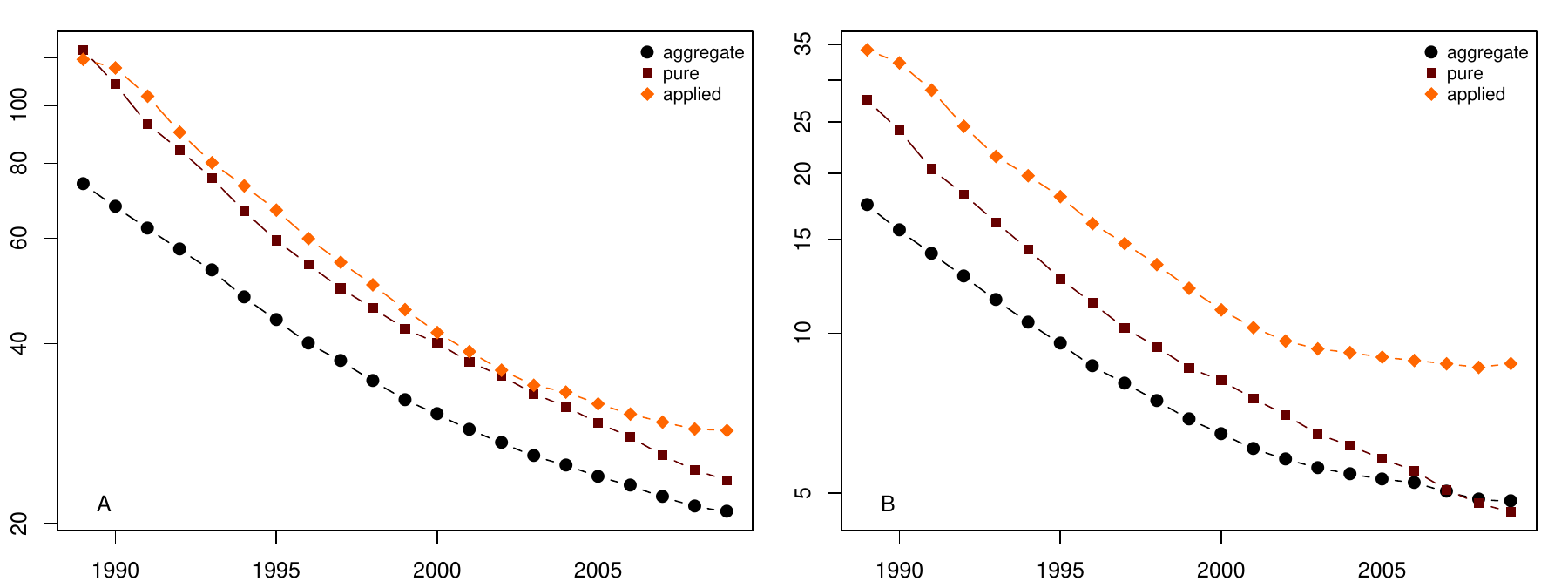}}
\caption{\small {\sc(a)} Harmonic mean distance and {\sc(b)} its normalization by (\ref{eqn:completecomponents}) across a 5-year sliding window.\label{fig:harmdist}}
\end{figure}

Fig.~\ref{fig:dist} {\sc(a)} depicts the raw average \(\overline{d}\) over time.
The arithmetic average shrank steadily in each of the aggregate, pure, and applied networks, from around 11 over 1985--9 to around 9 over 2005--9 in the aggregate.
The harmonic average, depicted in Fig.~\ref{fig:harmdist} {\sc(a)} (note the logarithmic vertical scale), decreased dramatically in contrast, from about \(74\) to about \(21\).
The adjacent boxplots {\sc(b)} depict the median (divider) and interquartile range (box) of each distribution.\footnote{Whiskers are omitted. When bound to the median by some small multiple of the interquartile range, the diameter in each case reduced the meaning of the whisker to precisely this bound; while whiskers allowed to extend to 1 and to the diameter in each interval crowd out the boxes for vertical space in the plot.}
Within each box are the arithmetic and harmonic averages.

Fig.~\ref{fig:dist} {\sc(c,d)} show that internode distances in \(G_1\) shrank less than one might expect due to rising density but kept pace with expectations based on the entire degree sequence.
The predictions themselves converged over time, with the empirical value sandwiched between them.
Fig.~\ref{fig:dist} {\sc(d)} suggests that this is an artifact of the changing degree sequence; the NSW model about matched \(G_1\) in the average internode distance, and in fact \(G_1\) gradually grew tighter-knit than the model.\footnote{Comparison to actual NSW models indicates that this is not an artifact of increased total size.}
Interestingly, the predictions themselves converged over time, with the empirical value situated between them.
This was almost entirely due to shrinking distances in the ER model (the distance distributions of NSW models were comparably steady in shape as well as in mean).

By normalizing the harmonic mean distance by components (Fig.~\ref{fig:harmdist} {\sc(b)}, again note the logarithmic scale), we see that the fragmentation of the network accounts for an order of magnitude's worth of the average distance; notably, accounting for components brought \(\overline{d^{-1}}^{-1}\) nearly into agreement with \(\overline{d}\).

Overall, \(G_1\) grew better connected over our 25-year interval in terms of internode distances than more basic connectivity indicators (density, degree sequence, and component size distribution) account for.
We attribute the sharp decline in \(\overline{d^{-1}}^{-1}\) to the changing distribution of component sizes, which as we saw had a huge impact on the calculation.
The different arithmetic but similar harmonic average distances in the pure and applied subnetworks may then be interpreted as reflecting a more fragmented applied network.
Indeed, when this is accounted for by the normalization of \(\overline{d^{-1}}^{-1}\), the applied appears {\it more} tightly-knit than the pure.
This in turn may be explained by the prevalence of highly-connected subcommunities in the applied network, often disconnected from the largest component.
This is suggested both by the smaller values of \(|C_1|\) in the applied network (Fig.~\ref{fig:comp}) and by the smaller sizes of the smaller components (not shown), and is consistent with the sensitivity of \(\overline{d^{-1}}^{-1}\) to short distances.

\subsection{Clustering}\label{clustering}

Short distances are half of the ``small world'' story; the other half is high clustering.
Clustering in graphs refers to the proliferation of triangles (pairwise linked triples):
The {\df (local) clustering coefficient} \(c_j\) of a research \(j\in G_1\) is defined to be the proportion of pairs of \(j\)'s collaborators who are themselvels collaborators \cite{ws-collective}.
The {\df (global) clustering coefficient} \(C\) of a graph itself is taken to be the proportion of triples \((j',j,j'')\) of any researcher \(j\in G_1\) and two of their collaborators \(j',j''\) that form triangles, i.e.\ for which \((j',j'')\in E(G_1)\) \cite{bw-properties}.
In social networks triangles far exceed expectations based on random graph models, and the sociological literature has explained this clustering in a variety of ways \cite{d-Davis,msc-birds}.

We measure clustering over time in three ways: the connectivity-dependent average clustering \(\overline{c}_k=\sum_{k_j=k}c_j/\sum_{k_j=k}1\) for \(k\geq 2\), the average clustering \(\overline{c}=\sum_{k_j\geq 2}c_j/\sum_{k_j\geq 2}1\), and the global clustering \(C\).
In addition to the raw numbers we consider the quotients of \(C\) and \(\overline{c}\) by the graph density (the expected level of clustering under the unipartite NSW model) and the quotient of \(C\) by its expected value \(C_{\rm bNSW}\) under the bNSW model, computed in \cite{nsw-random} as\begin{equation}\label{eqn:nswclustering}
C_{\rm bNSW}\equiv\bigg(\frac{(\mu_2-\mu_1)(\nu_2-\nu_1)^2}{\mu_1\nu_1(2\nu_1-3\nu_2+\nu_3)}+1\bigg)^{-1}\text,
\end{equation}
where \(\mu_r=\sum_j{q_j}^r\) and \(\nu_r=\sum_i{a_i}^r\) are the \(r^\text{th}\) moments of the distributions of researcher productivity and of publication cooperativity, respectively.\footnote{The bipartite model predicts different connectivity distributions than what we observe in \(G_1\), so degree-dependent comparisons to this model would require somewhat deeper discussion.}
Comparisons to ER will indicate the level of clustering relative to the baseline given by graph density, or average connectivity; comparisons to NSW will indicate what clustering that cannot be accounted for by the cooperativity of publications alone.\footnote{It is possible for measured clustering to be lower than that predicted by the NSW model, as in \cite{nsw-random} (company directors), should very little clustering be due to distinct pairwise collaborations and many highly cooperative publications share a common pool of authors, which publications would in the model be attributed to distinct teams of researchers.}

Global clustering in coauthorship graphs ranges widely, across \(.066<C<.76\) over intervals of time close to ours (5 years), but higher clustering is far more common \cite{bjnrsv-evolution,g-evolution,n-scientific-I}.
Adopting our interpretation of the nodes, clustering in bipartite projections like \(G_1\) occurs when three (or more) researchers coauthor a publication and when each pair of a triple of researchers has coauthored something without the other.
The respective explanatory power of these process has received limited attention \cite{gl-bipartite,o-triadic}.
In such cases, the measured ratios of \(C_{\rm bNSW}\) to \(C\) were similar, .42 for the {\tt arXiv} and .48 for MEDLINE \cite{nsw-random}.

\begin{figure}[h]
\centerline{\includegraphics[width=\columnwidth]{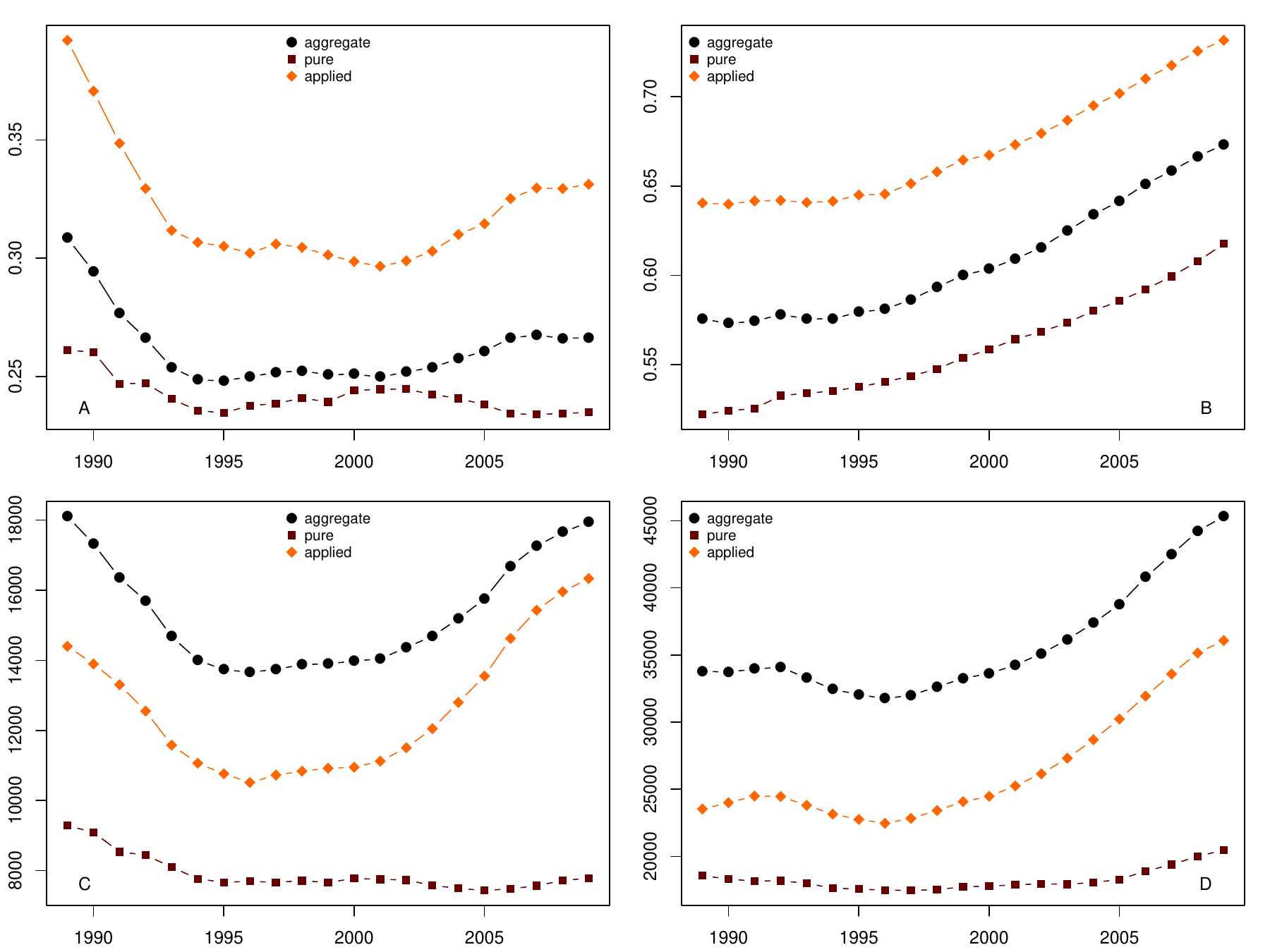}}
\caption{\small Across a 5-year sliding window:
{\sc(a)} global clustering coefficient \(C\),
{\sc(b)} average local clustering coefficient \(\overline{c}\), and
the ratios {\sc(c)} of \(C\) and {\sc(d)} of \(\overline{c}\) to \(2m/n(n-1)\).\label{fig:clust}}
\end{figure}

Fig.~\ref{fig:clust} {\sc(a,b)} depict the global and average local clustering coefficients over time.
Clustering in \(G_1\) was lower than typical for collaboration networks, in the range \(.24<C<.31\), with the applied network exhibiting consistently higher levels than the pure.
Whereas \(C\) decreases until 1990--4, after which it stabilizes, \(\overline{c}\) had been steady until this time and then began to rise.
Since the local average is more sensitive to the high local clustering \(c_j\) of researchers with low connectivity \(k_j\), this coincidence may be explained by the changing distribution of connectivity around the same time (see the discussion of Fig.~\ref{fig:deg} {\sc(c)}) amidst a more or less steady rise in clustering across researchers.
Time series of \(\overline{c}_k\) across \(2\leq k\leq 12\) (see the supplementary materials) show that the earlier period (before the mid-90s event) was characterized by consistently rising clustering only among low-connectivity researchers, while the later period saw a more rapid increase across researchers of all connectivities.

\begin{figure}[h]
\centerline{\includegraphics[width=\columnwidth]{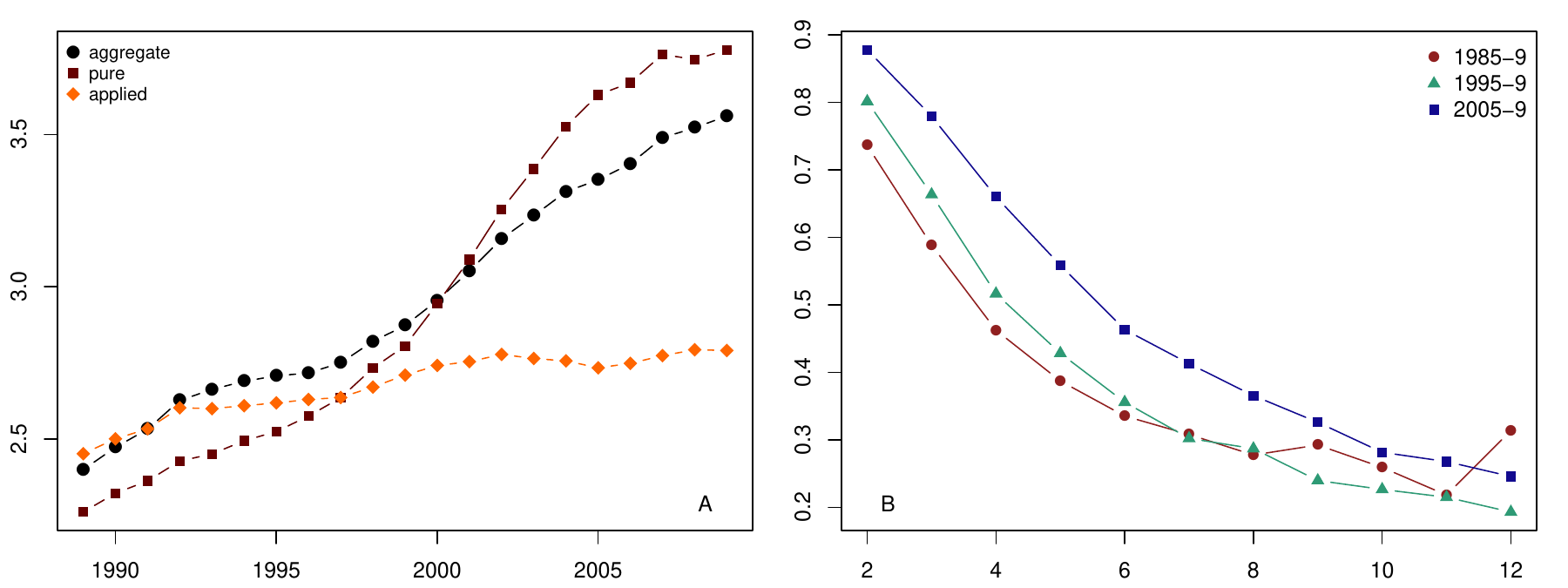}}
\caption{\small {\sc(a)} The ratio of \(C\) to (\ref{eqn:nswclustering}) across a 5-year sliding window and {\sc(b)} average connectivity-dependent clustering coefficient \(\overline{c}_k\) versus \(k\).\label{fig:clustalt}}
\end{figure}

Fig.~\ref{fig:clust} {\sc(c,d)} and \ref{fig:clustalt} {\sc(a)} show these clustering coefficients normalized by model predictions.
The density of \(G_1\) accounted for little of the long-term trends in clustering, as the time series are only slightly distinguishable.
Notice the higher ratio of \(C\) and \(\overline{c}\) to density in the aggregate.
The lower density of the aggregate network than the pure or applied separately, which also played into the higher levels of intra- than inter-disciplinary links in \(G'_1\), accounts for this.
Cooperativity, on the other hand, accounted for between 28\% and 42\% of the observed clustering in the aggregate, at first about as much as in previously-studied collaboration networks but less as time progressed.

Clustering trends in the two major networks relied on different phenomena.
The comparison of Fig.~\ref{fig:clust} {\sc(b)} with {\sc(d)} suggests that increased local clustering in the pure network was adequately explained by rising average connectivity, while the comparison of Fig.~\ref{fig:clust} {\sc(a)} with Fig.~\ref{fig:clustalt} {\sc(a)} suggests that changes in global clustering in the applied network was largely due to the proliferation of highly cooperative publications.

\subsection{Trends across disciplines and over time}\label{differences}

We have discerned several differences between the pure and applied subnetworks, and between the evolutionary trends over the periods within our 25-year interval loosely defined by the two major events.
While we do not conduct a thorough analysis of these differences, we take a preliminary look in terms of widely-used network diagnostics.

Differences in publishing culture and in external influences may have a strong impact on the respective structures of the pure and applied networks (see Section~\ref{discussion}).
However, it is worth considering first the possible impact of the MR demarcation of the literature itself.
Whereas most of the collaboration conducted by more pure mathematicians is likely to be with other mathematicians, applied mathematicians are more likely to collaborate with non-mathematicians.
This leads us expect (a) that pure mathematics and its researchers are situated more centrally in the MR network, with applied mathematics and its researchers more toward the periphery; and (b) that applied mathematicians form a less cohesive network than pure.
The expectation (b) is supported by the greater fragmentation of the applied network in terms of its smaller largest component, larger internode distances within that component, and greater fragmentation among components, observed in Sections~\ref{connectedness} and~\ref{distance}.

The expectation (a) may be tested in terms of the pure versus applied research interests of the researchers that appear more centrally in \(G_1\).
In particular, we might expect that researchers of greater betweenness, closeness, and eigenvalue centrality---properties influenced by nodes' positions within the entire network---should tend to have authored more pure publications, in contrast to researchers of greater degree or weighted degree centrality---properties that are strictly local \cite{wf-social}.
We therefore consider, as \(x\) ranges from 1 to 1000, the attributions among the \(x\) most central researchers that are pure, as a proportion of those that are pure or applied (according to primary MSC).
We do this over three evenly-spaced 5-year windows for degree, weighted degree, closeness, betweenness, and eigenvalue centrality.

\begin{figure}
\centerline{\includegraphics[width=\columnwidth]{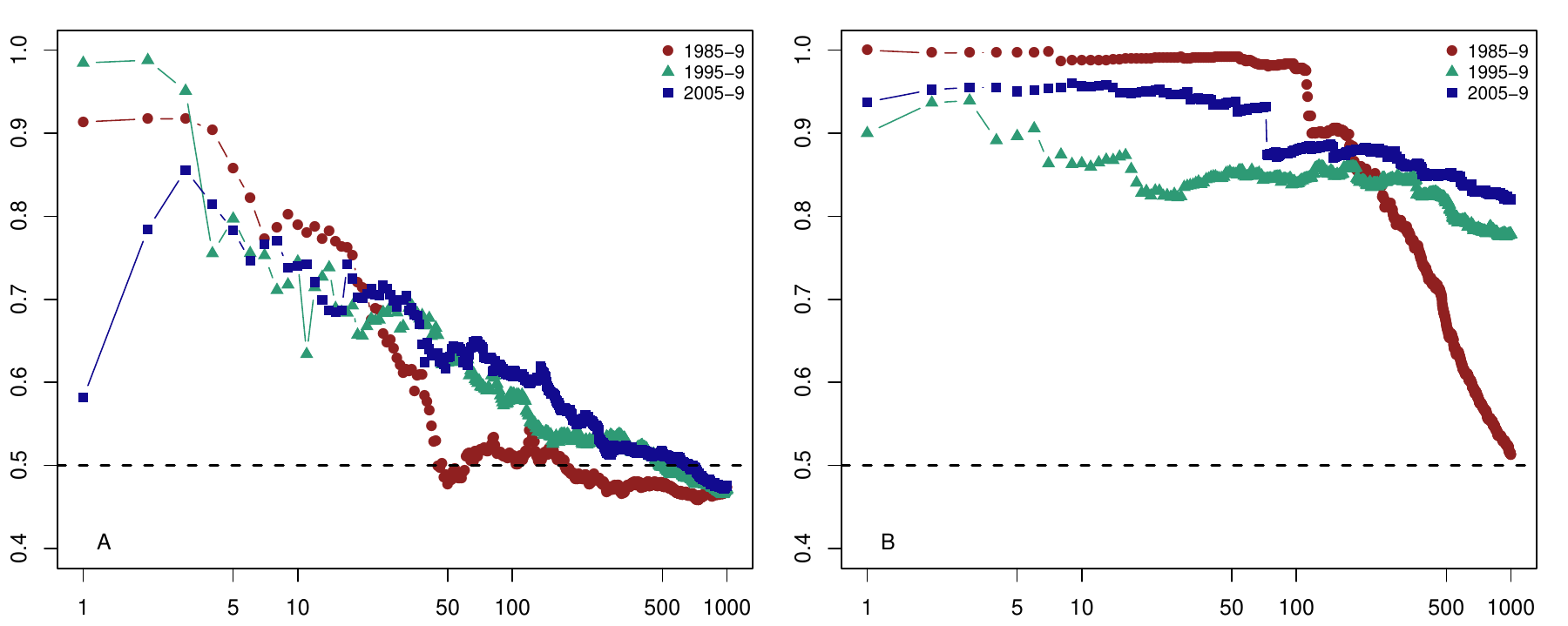}}
\caption{\small Proportion of pure and applied attributions to {\sc(a)} the \(x^\text{th}\) highest betweenness and {\sc(b)} the \(x^\text{th}\) highest eigenvector centrality researchers that are pure across \(1\leq x\leq 1000\) over three 5-year windows.\label{fig:cent}}
\end{figure}

The results for betweenness and eigenvalue centrality are depicted in Fig.~\ref{fig:cent}; those for degree, weighted degree, and closeness were similar in shape to those for betweenness.
Consistently over time and across centrality measures, researcher attributions began disproportionately pure.
In all but eigenvalue centrality, they declined rather steadily toward a more balanced proportion by the time the top 100 or so researchers had been included.
While the similarity of closeness and betweenness centrality trends to those of (weighted) degree is dissuasive of the idea that pure researchers occupy the ``center'' of the MR network, the persistently disproportionately pure research focus of high--eigenvalue centrality researchers suggests that, in terms of structural ``influence'' or ``importance'', pure researchers are indeed central to the discipline as a whole.

We have until now discussed changing trends in network evolution as though the mid-90s and early-00s events were common, coordinated phenomena being felt by a variety of network diagnostics.
While some of these trends are certainly related (trends in cooperativity and connectivity, for example), there is an alternative hypothesis that multiple network trends, not directly interrelated, have been approximately coincident.
This is suggested by the apparent changes in trend of multidisciplinarity \(\overline{s}\), which are more numerous than and not coincident with the two events, as we have described them.
We undertake now to (1) see just how coincident were the fluctuations we observed; (2) assuming that they were, glean the order in which they proceeded; and (3) glean how sensitive the answers to both are to some of the most impactful researchers and publications.

To get a handle on when each time series changed course, we use a type of {\df change point model} \cite{ka-change-point,p-continuous}.
Specifically, to the ordered pairs \((t,D(G_t))\) we fit the continuous, piecewise-linear model
\[D(G_t)=\beta_{0}+\beta_{1}t+\beta_{2}(t-c)\delta_{t>c}+\epsilon_t,\ \ \ \ 1984+\Delta t\leq t\leq 2009\text,\]
having normally distributed error \(\epsilon\).\footnote{Our code in {\tt R} uses the {\tt nls} function to locate maximum-likelihood estimators, i.e.\ those that minimize \(SSE=\sum_t{\epsilon_t}^2\).}
There is some subjectivity in how the algorithm is initiated and in how the windows surrounding each change point are chosen, and moreover it is not necessarily likely that the network evolves in a piecewise linear fashion.
The models fit the data reasonably well, however, and we take advantage of them only locally, to situate abrupt shifts in the data relative to each other in time.
That is, if the shift in diagnostic \(D\) occurred before that of diagnostic \(D'\), then we expect the estimated change point \(\hat c\) from the fit to \((D(G_t))\) to be smaller than that from the fit to \((D'(G_t))\).
We exhibit code and all change point fits to time series in the supplementary materials.

\begin{figure}[h]
\centerline{\includegraphics[width=\columnwidth]{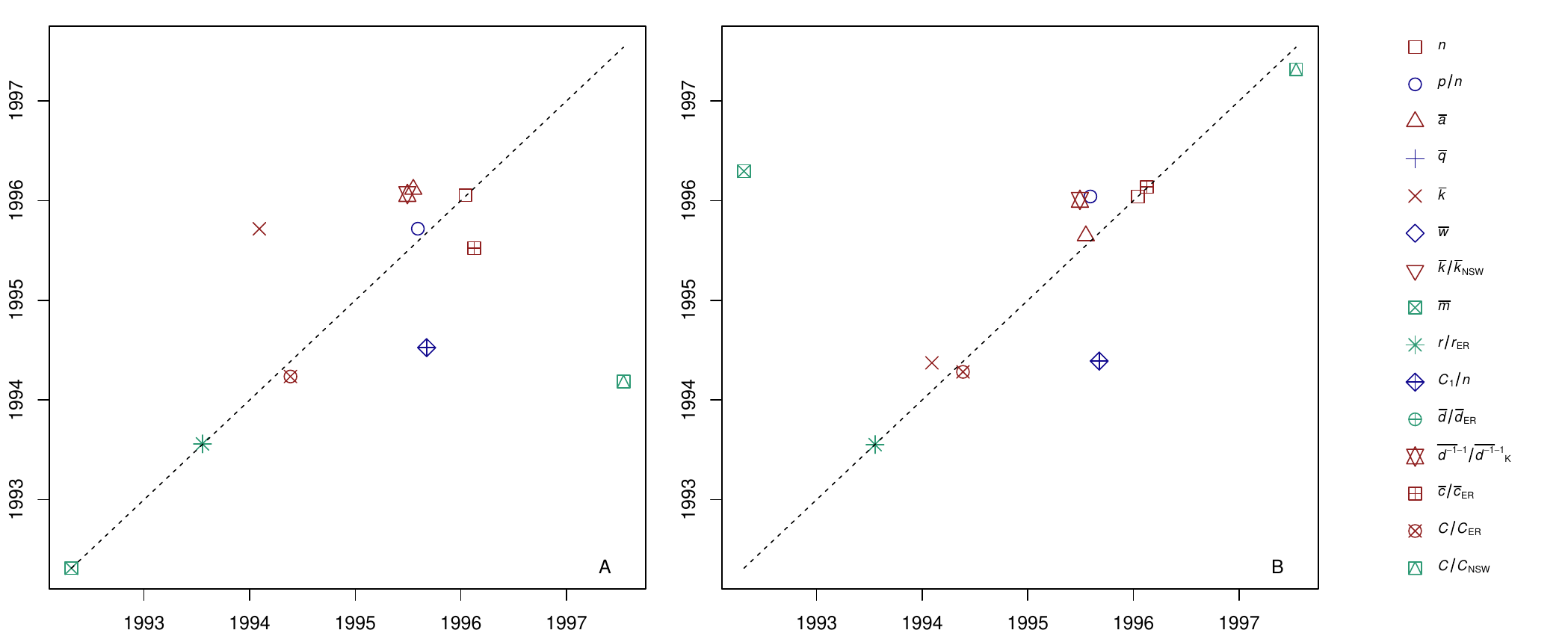}}
\caption{\small Delay plots for the first change point.
Each point in each plot has \(x\)-coordinate the best-fit change point to the time series of a diagnostic on the aggregate network and \(y\)-coordinate the same on the alternative network labeled on the vertical axis.
Some ordered pairs are not plotted because change points were not estimated.
Each dotted line depicts the relationship \(y=x\).\label{fig:delay}}
\end{figure}

The time series we use for this analysis are listed in the legend and caption to Fig.~\ref{fig:delay}.
These were chosen from among the time series discussed up to this point, with preference given to those of very basic diagnostics (for instance, network size and average cooperativity) and to those of other diagnostics (for instance, number of publications and global clustering), divided by their expectations based on more basic ones (number of researchers and bipartite NSW model, respectively).
We performed a correlational analysis on the 15 time series chosen, the results of which we include in the supplementary materials.
Based on this analysis, we sorted the time series into three groups: a largest group that were tightly correlated, a smaller group that were moderately correlated with each other and with the larger group, and a smaller group that were tightly correlated with each other but negatively correlated with the others.
These groups are identified in Fig.~\ref{fig:delay} by the colors red, green, and blue, respectively.

In order to test the sensitivity of these observations to highly influential researchers and publications, we perform the same analysis on a ``few-author'' network constructed from those publications \(i\) having cooperativity \(a_i<7\) and a ``less prolific'' network obtained by removing (for each 5-year window) those researchers \(j\) having productivity \(q_j\geq 48\).\footnote{The threshold for cooperativity is chosen to be the values for which publication counts {\it decreased} until the mid-90s. The other two thresholds are chosen so that the proportion of researchers removed to obtain the second and third alternatives most closely resembles the proportion of publications removed to obtain the first.}
To account for the possible influence of window size, we repeat the process across a 3-year sliding window.
The results are essentially the same; see the supplementary materials for details.

\begin{figure}[h]
\centerline{\includegraphics[width=\columnwidth]{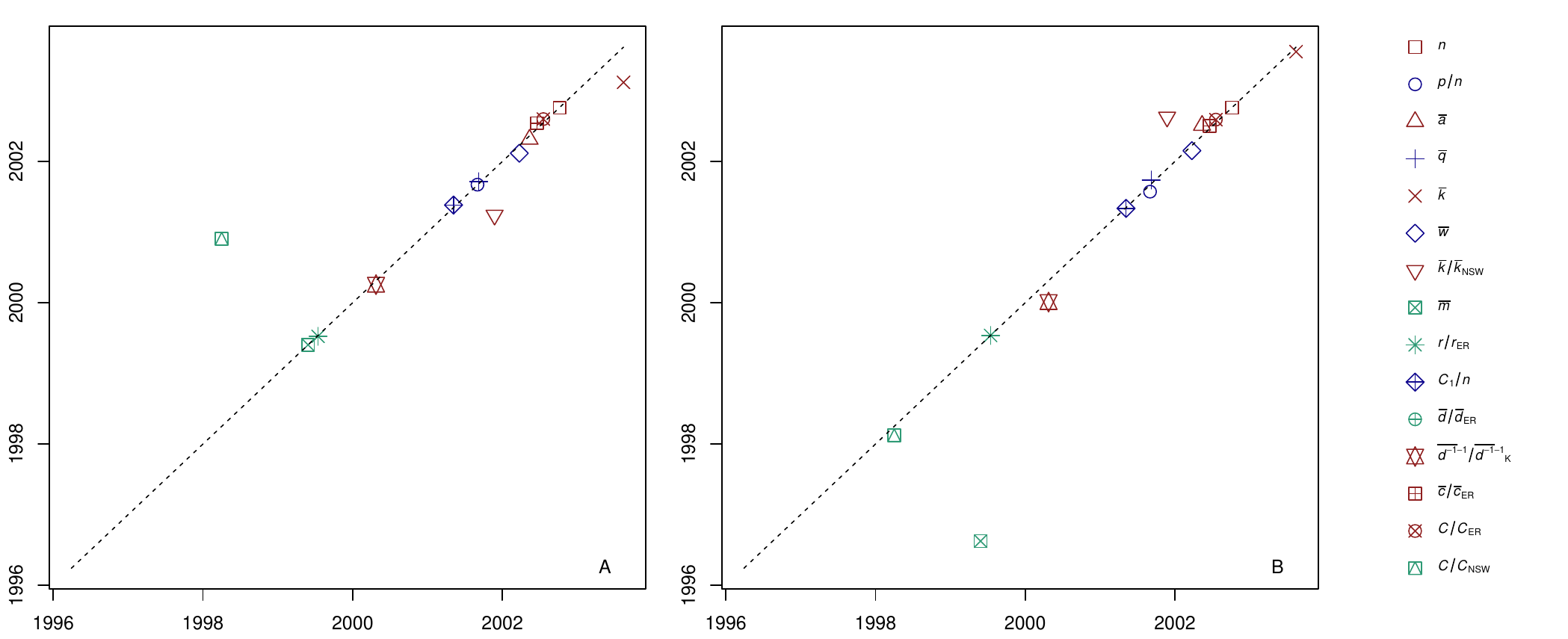}}
\caption{\small Delay plots for the second change point.\label{fig:delay2}}
\end{figure}

Fig.~\ref{fig:delay} depicts ``delay plots'' that record, for each event and for each pairing of the aggregate network with one of the aforedescribed alternatives, the estimates of the change point \(c\) close to the event on the time series of several network diagnostics described in earlier sections.
We make two main observations:
First, change points for measures of cross-disciplinarity and clustering vary more widely, both among each other and between the aggregate and alternative networks, than those for measures of connectedness, output, and cohesion.
This likely has to do with the latter being mostly averages across nodes, which would be less sensitive to the removal of top players than global diagnostics.
This possibility is supported by the observation that the normalized average local clustering \(\overline{c}\) (the upward-pointing open pink triangle) behaves more like the latter group than like the former.\footnote{While we do not include it here, degree-degree correlations, measured as assortativity \cite{n-mixing}, varies similarly.}
Second, change points corresponding to the mid-90s event vary more widely, in the same ways, than those corresponding to the early-00s event.
That is, the time series shifts around this time were more coincident.
While these plots are suggestive, the reader should bear in mind that they do not take into account the suitability of the change point model (considered in the supplementary materials).

\section{Discussion}\label{discussion}

We observe several consistent trends  in the long-term evolution of the MR collaboration network:
Both the research community and the published literature grew at increasing rates, and the community decidedly more so.
These trends are largely explained by greater cooperativity in publishing (papers having three or more authors) and greater connectivity among researchers (those having three or more collaborators), including proportional declines in solo publications and solo researchers.
In particular, increasingly many of the authors of the most cooperative publications publish little else (in mathematics).
Meanwhile, the network has grown better-connected even than this increased connectivity suggests:
Internode distances grew steadily shorter than random graph models having the same density, connectivity distribution, or size distribution of connected components predicted.
Simultaneously, clustering steadily increased, both at the local level and at the global level, and especially clearly once clustering due to cooperativity was taken into account.

These trends and their discrepancies might be interpreted in several, compatible ways.
It may be that as researchers become better-connected more avenues emerge for collaborative projects, resulting in a literature more dense with contributions per paper overall.
This hypothesis is supported indirectly by the steady increase in researcher clustering but countered directly by a weak relationship between cooperativity and multidisciplinarity.
Alternatively, whereas the enlarged community includes many researchers who publish very little, we may be detecting the involvement of researchers who are not career mathematicians (or at any rate whose career research is not covered by MR) but who join mathematics research teams only once or infrequently.
These would include peers in other fields and young researchers who progress on to other fields after a program in mathematics.
A reciprocal trend should therefore also be observable as an increase in infrequent authorship by researchers in collaboration networks of other disciplines that collaborate often with mathematics.
It also suggests a third possible explanation: that the overall scientific literature is itself becoming a more cohesive network, in the same way as the pure and applied networks are growing more cohesive within mathematics.
This should be observable as a general trend across all collaboration networks toward increasing community size relative to the literature.
This hypothesis also implies an upward trend in the proportion of common-author ties between pure and applied publications (links in \(G'_1\)), relative to all such ties---which we observe until the mid-90s event and after the early-00s event.
Only during the latter period did the research community show exceptional growth, as visible in Fig.~\ref{fig:size} {\sc(b)}.
These explanations may amount to the common phenomenon of increased cohesion throughout scientific publishing being observed at different scales.

The partition of the literature into ``pure'' and ``applied'' based on primary subject classification yields two literatures of very nearly equal size, which together comprise more than 97\% of the aggregate literature over any 5-year interval.
The research communities, while they overlap substantially, are also approximately balanced in number until the mid-90s event and remain close.
Other long-term trends in both subnetworks mimicked those in the aggregate.
Despite these similarities, the networks exhibited some interesting differences, most of which persisted over our 25-year interval and hence suggest essential differences between the literatures.
The surge in one-time authors and in one-time collaborations were concentrated in the applied subnetwork, which also exhibited greater connectivity, more short distances, and higher clustering.
The pure subnetwork showed greater productivity overall, in terms of individual researchers and of collaborating pairs.

Moreover, pure research was consistently more multidisciplinary, as measured by the number of assigned subject classifications.
This difference may reflect the scope of the database; it should be expected if {\tt MathSciNet} records a great deal of interdisciplinary work among different branches of mathematics but only a subset of interdisciplinary work among mathematics and other disciplines (much of which would be published in non-mathematics journals).
Alternatively, it may reflect greater frequency of collaboration among mathematicians in different subfields than among mathematicians and other researchers.
An analysis of a more general scientific publishing database, with a comprehensive inter- and intra-disciplinary classification scheme, could lend support to one of these options over the other.

Both subnetworks showed increased clustering and decreased distances over our interval, suggesting an ongoing ``small world'' effect that also manifests in the aggregate.
Interestingly, while the pure network exhibits shorter distances, the applied exhibits higher clustering.
Neither, therefore, may be said to be the ``superior'' small world.
It is tempting to interpret this as an illustration of the trade-off between low distances and high clustering.

However, each observation can be understood in terms of more basic phenomena.
The shortening of distances in both (and the aggregate) can be adequately accounted for by the sheer increase in connectivity or density (see Fig.~\ref{fig:dist} {\sc(c)}), while much of the rise in clustering, especially in the applied subnetwork, was due to the proliferation of several-author publications (see Fig.~\ref{fig:clustalt} {\sc(a)}).
The shorter distances in the applied network are largely due to the researchers who publish papers in large groups, and especially those who are removed from the largest component (Fig.~\ref{fig:harmdist}), and once cooperative publications are taken into account only the pure network shows a steady rise in clustering (Fig.~\ref{fig:clustalt} {\sc(a)}).

It may also be that the pure and applied networks are situated differently within the MR literature in such a way as to produce some of these differences as artifacts.
We suggest, for instance, that the pure network may feature more centrally in this data, which could account for its higher productivity and greater cohesion (into a largest component), whereas the applied occupies more of the periphery, where one-time authors surged in the last decade.
The proportion of pure versus applied research contributions among the most central researchers is suggestive of this, especially that the researchers of greatest eigenvalue centrality have authored nearly uniformly pure research.
More sophisticated structural measurements and models, or comparisons to other databases, would be needed to more carefully answer this question.

Changing rates of growth in the network are noticeable but perhaps not suspicious.
We found that these fluctuations in growth (both in community and in output) are not just quantitative; they occur simultaneously with dramatic changes in network structure and may need to be understood in terms of many factors.

The two events, such as we have described them, tell dissimilar stories.
The mid-90s event was characterized by noticeable increases in the rates at which the research community and literature grew.
This growth was coupled with a trend toward greater local connectivity and clustering, especially among applied researchers.
The event also saw brief declines in cross-disciplinarity.
Thought of as a single phenomenon, the event took place over several years and was significantly influenced by the rise of highly cooperative publications.
Meanwhile, the early-00s event was characterized by decreased individual and collaborative publishing rates, due in large part to an influx of few-time authors.
A surge in several-author publications, to which many few-time authors contributed, wrought a surge in clustering, again especially in the applied subnetwork.
Though connectivity continued to increase on average, following this event it was to a lesser extent than the random bipartite model would predict, and by other diagnostics (largest component and internode distances) the increasing cohesion of the network slowed.
This was a more coordinated event, in that shifts in time series were more coincident (see Fig.~\ref{fig:delay} {\sc(c,d)}), and less sensitive to the contributions of specific publications or researchers.

The mid-90s event may have been due in part to several plausible factors.
One was the rise of e-communications and the World Wide Web:
Among the Internet milestones that have impacted academia are the introductions of the {\tt arXiv} in 1991, which went online in 1993 \cite{ginsparg-global}, and of {\tt MathSciNet} in 1996, which made the MR publishing database available through a graphical web interface \cite{j-Chinese}.
Another was the influx of mathematicians from the former Soviet Union into the MR database, whether by moving to other institutions or by their research becoming accessible to MR \cite{bd-collapse}.
While the early-00s event was more precisely situated in time, and more clearly the result of specific publishing trends, we don't feel prepared to speculate on its likely proximate causes or on whether its impact is likely to have been beneficial for mathematical research on the whole.

\section{Conclusions}\label{conclusion}

While evolving and temporal models and diagnostics of time-resolved network data are seeing widespread use, the changing structure of collaboration networks with respect to traditional diagnostics and model predictions has not been widely studied for its own sake.
In particular, few evolving networks have been studied with an eye toward examining abrupt changes in their evolution, and evolutionary and temporal models are generally designed rather to reproduce steady behaviors than to account for such changes.
We examined the collaboration network of mathematicians as constructed from the MR database over the period 1985--2009 with the aim of understanding what essential trends describe the network's evolution, how the network structure differs by discipline, and in particular how network evolution deviates from long-term trends and what factors may account for such behaviors.

Several trends were straightforward over this period, including increasingly rapid growth in both the community and in its output and greater connectedness (over a fixed period of time).
The latter is indicated in several different ways: larger teams of coauthors, greater total numbers of collaborators per researcher, greater proportions of researchers connected through coauthorship, shorter distances through coauthorship between researchers, and increased collaboration among a typical researcher's collaborators.
Moreover, these trends were not explainable in terms of each other; graph models tailored to mimic the network according to some of these trends, but to otherwise exhibit random structure, do not account for others.
It is fair to say by any standard that the network has grown better-connected.
(The literature also showed some signs of an increased multidisciplinary quality, but this deserves closer scrutiny.)

Two major subnetworks, loosely corresponding to more pure and to more applied disciplines within mathematics, exhibit similar long-term trends and fluctuations to the aggregate.
They also exhibit significant and consistent structural differences with each other.
These may be explained in terms of how disciplines are situated within the larger network, of how mathematicians in different specialties engage with other researchers, or of different research cultures within mathematics.
Several questions could be asked and answered graph-theoretically to tease these and other explanations apart.

Steady trends in the evolution of the MR network divide our 25-year interval into three segments, separated at two moments we call events: one in the mid-1990s, the other in the early 2000s.
Both events heralded growth and greater connectivity in each of the networks we studied (aggregate, pure, and applied), but on closer inspection they show important differences.
We speculated that several real-world phenomena may have factored into these events.
Closer study of the MR database could provide greater insights, and evidence from other sources could better inform and discriminate among these hypotheses.

\begin{acknowledgements}
The authors are grateful to the American Mathematical Society for providing access to the {\it MR} database and for agreeing to make the data publicly available (by request to the Executive Director).
The authors thank Sastry Pantula and Philippe Tondeur for helpful information, and Sid Redner, Betsy Williams, and participants of the Summer 2010 REU in Modeling and Simulation in Systems Biology for helpful conversations and support.
\end{acknowledgements}

\bibliographystyle{spmpsci}      
\bibliography{MRtext}   

%
%

\end{document}


\title{Evolutionary Events in a Mathematical Sciences Research Collaboration Network
}


\author{Jason Cory Brunson \and Steve Fassino \and Antonio McInnes \and Monisha Narayan \and Brianna Richardson \and Christopher Franck \and Patrick Ion \and Reinhard Laubenbacher
}


\institute{J. C. Brunson \at
              Virginia Bioinformatics Institute,
              Washington St, MC 0477, Virginia Tech, Blacksburg, VA  24061 \\
           \and
           S. Fassino \at
              Department of Mathematics,
              227 Ayres Hall, University of Tennessee, Knoxville, TN  37996
           \and
           A. McInnes \at
              Department of Mathematics and Computer Science,
              Oakwood University, Cooper Complex Bld. B, 7000 Adventist Blvd, Huntsville, AL  35896
           \and
           M. Narayan \at
              Lyman Briggs College,
              Michigan State University, 35 East Holmes Hall, East Lansing, MI  48825
           \and
           B. Richardson \at
              Department of Mathematics and Computer Science,
              Oakwood University, Cooper Complex Bld. B, 7000 Adventist Blvd, Huntsville, AL  35896
           \and
           C. Franck \at
              Laboratory for Interdisciplinary Statistical Analysis,
              212 Hutcheson Hall, Blacksburg, VA  24061
           \and
           P. Ion \at
              Mathematical Reviews,
              P.O. Box 8604, Ann Arbor, MI  48107
           \and
           R. Laubenbacher \at
              Center for Quantitative Medicine,
              University of Connecticut Health Center,
              195 Farmington Ave,
              Farmington, CT  06030
              Tel.: +123-45-678910\\ 
              Fax: +123-45-678910\\
              \email{laubenbacher@uchc.edu}           
}

\date{Received: date / Accepted: date}


\section{Supporting Information}

\subsection{Cooperativity--multidisciplinarity relationship}

\begin{figure}[h]
\centerline{\includegraphics[width=\columnwidth]{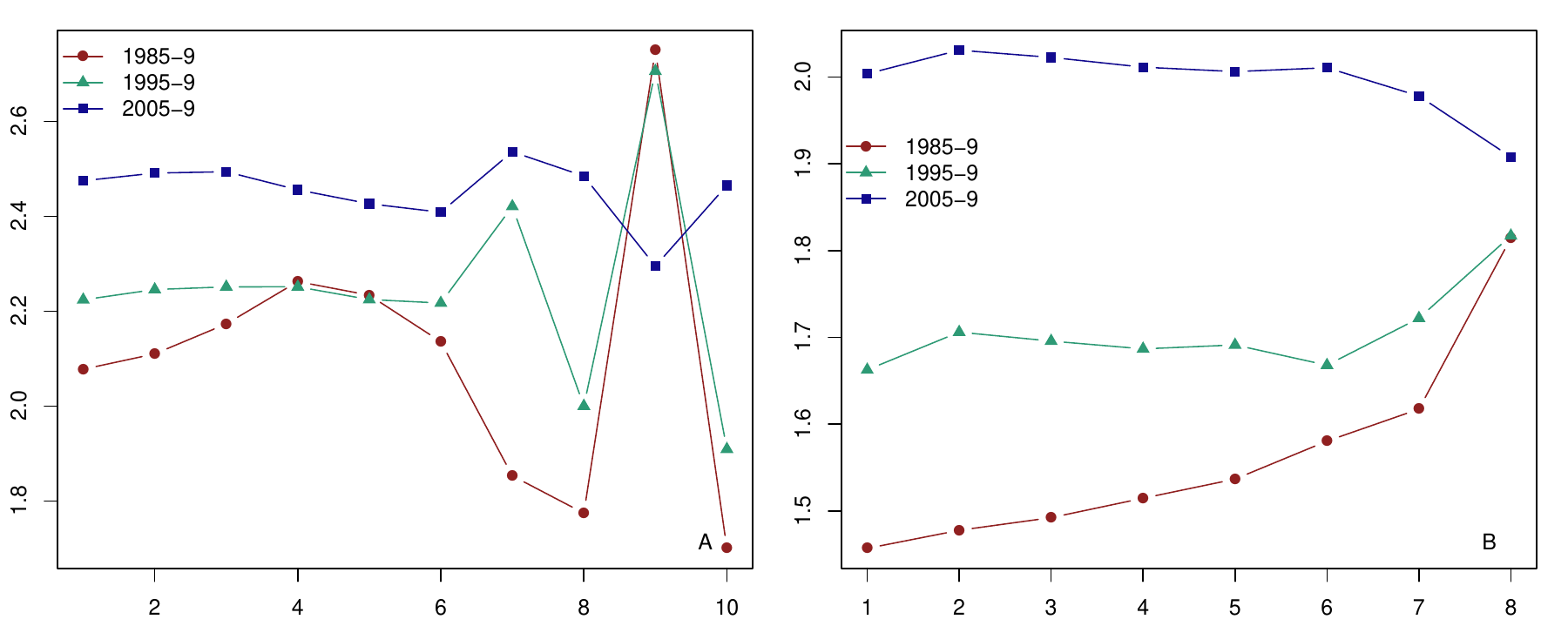}}
\caption{\small For three evenly-spaced 5-year intervals,
{\sc(a)} for values \(a=1,\ldots,8\), the expected mean multidisciplinarity of the publications \(i\) with cooperatively \(a_i=a\), and
{\sc(b)} for values \(m=1,\ldots,10\), the expected mean cooperativity of the publications \(i\) with productivity \(m_i=m\).\label{fig:am}}
\end{figure}

Fig.~\ref{fig:am} depicts
\begin{equation}\label{eqn:multidisciplinarity-vs-cooperativity}
\frac{\sum_{m_i=m}a_i}{\sum_{m_i=m}1}\ \text{versus}\ m
\hspace{.25cm}\text{and}\hspace{.25cm}
\frac{\sum_{a_i=a}m_i}{\sum_{a_i=a}a_i}\ \text{versus}\ a
\end{equation}
across a 5-year sliding window.

\subsection{Connectivity-dependent clustering}

\begin{figure}[h]
\centerline{\includegraphics[width=.5\columnwidth]{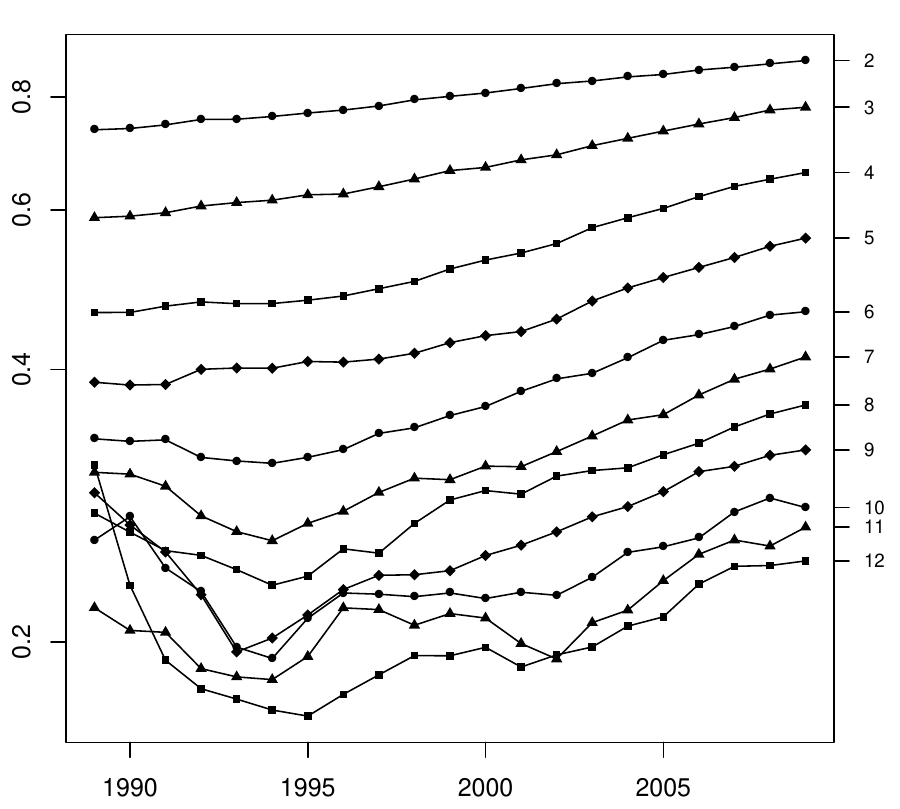}}
\caption{\small Across a 5-year sliding window, \(\overline{c}_k\) for \(2\leq k\leq 12\).
Each time series \(\overline{c}_k\) is identified by its value of \(k\) on the right vertical axis.\label{fig:connclust}}
\end{figure}

Fig.~\ref{fig:connclust} depicts connectivity-dependent clustering \(\overline{c}_k=\sum_{k_j=k}c_j/\sum_{k_j=k}1\) for connectivities \(k=2,\ldots,12\) (note the logarithmic scale of the vertical axis).
It is immediately clear how the time series for average clustering depicted in Fig.~10 {\sc(b)} is largely determined by the (far more populous) researchers of low connectivity and high clustering, while that for global clustering in {\sc(a)} is determined by the researchers of higher connectivity around whom far more triangles congregate.
In particular, researchers having 12 collaborators were remarkably highly clustered early in our 25-year interval before declining rapidly; following the mid-90s event they exhibit clustering in keeping with the widely-observed negative relationship between \(k\) and \(\overline{c}_k\).

\subsection{Change point models}

We fit change point models to time series data that were not clearly piecewise linear, but that exhibited one or two major changes in behavior amid smaller perturbations that the models interpret as normally-distributed error.
The principle behind change point models is the same as that behind linear models.
A traditional linear fit takes the form
\[y_i=\beta_0 + \beta_1 x_i + \epsilon_i,\ \ \epsilon_i\overset{\rm iid}{\sim}N(0,\sigma)\text,\]
while our change point model takes the form
\[y_i=\beta_{0}+\beta_{1}x_i+\beta_{2}(x_i-c)\delta_{x_i>c}+\epsilon_i,\ \ \epsilon_i\overset{\rm iid}{\sim} N(0,\sigma)\text.\]
We caution that some basic statistical assumptions for change point models are not met by this data:
Particularly because adjacent sliding windows share 4 out of their 5 years but also because most authors publish in multiple years, measurements performed on these windows cannot be considered independent.
Because each window contains a different (increasing) number of publications, they cannot be considered identically distributed.
By performing change point analysis we do not intend to make predictions of future behavior but only to take advantage of an effective method for identifying shifts in behavior otherwise well-modeled linearly.
All data and computations are available upon request to the first author.

\begin{figure*}[h]
\centerline{\includegraphics[width=\columnwidth]{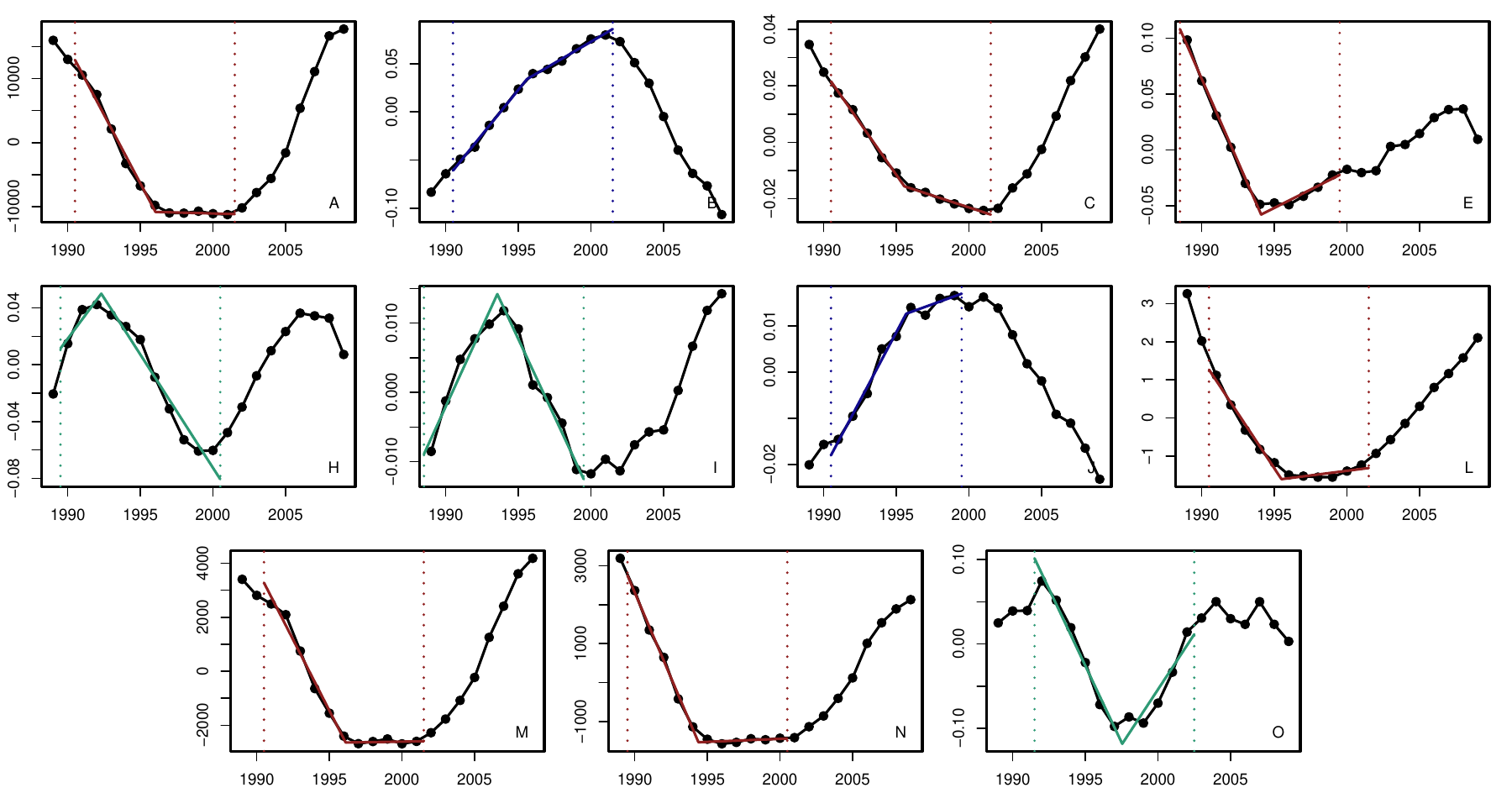}}
\caption{Residuals from best linear fits overlaid with a change point fit about the mid-90s event (5-year sliding windows).\label{fig:cp1}}
\end{figure*}

What follows is a simplification of the code we used to perform change point analysis in {\tt R}.
We required a guess {\tt c} at the change point and calculated estimators for the coefficients by fitting a linear model {\tt lm1} to the data below {\tt c} (providing \(\hat\beta_0\) and \(\hat\beta_1\)) and a linear model {\tt lm2} with fixed intercept at \(({\tt c},{\tt lm1}({\tt c}))\) to the data above {\tt c} (providing \(\hat\beta_2\)).

\begin{verbatim}
# FUNCTION: Change point analysis on a collection of ordered pairs
changepoint.model <- function(
  x,    # points (independent), sorted
  y,    # values (dependent)
  c     # number in range(x)
) {
  len <- length(x)
  stopifnot(len == length(y))
  # Linear model to estimate b0 and b1
  m <- max(which(x < c))
  lm1 <- lm(y[1:m] ~ x[1:m])
  b0 <- lm1$coeff[1]
  b1 <- lm1$coeff[2]
  # Scaling model to estimate y-value of b2 at x = c
  int <- lm1$coeff[1] + lm1$coeff[2] * c
  # x-values with origin (c,int)
  x2 <- x[(m + 1):len] - c
  # y-values with origin (c,int)
  y2 <- y[(m + 1):len] - int
  lm2 <- lm(y2 ~ x2 + 0)
  b2 <- lm2$coeff[1] - lm1$coeff[2]
  # Change point model using estimators for c (given), b0, b1, and b2
  return(summary(nls(
    as.formula('y ~ B0 + B1 * x + B2 * (x - C) * (x >= C)'),
    start = list(C = c, B0 = b0, B1 = b1, B2 = b2)
    )))}
\end{verbatim}

Fig.~\ref{fig:cp1} depicts the change point fits we used to examine the two events in the main paper.
In each plot the dotted vertical lines demarcate the intervals used to construct the model.

\begin{figure*}
\centerline{\includegraphics[width=\columnwidth]{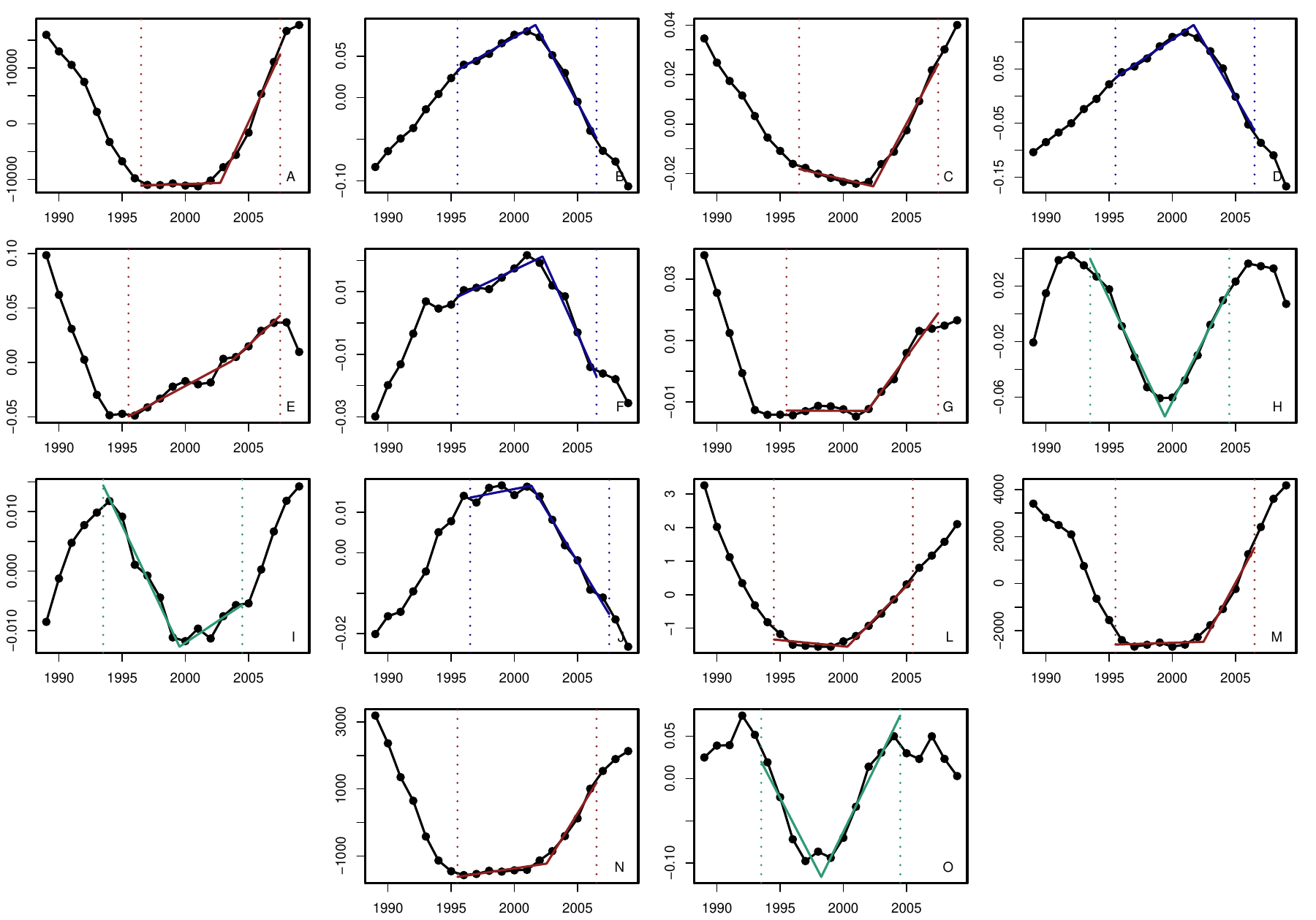}}
\caption{Residuals from best linear fits overlaid with a change point fit about the early-00s event (5-year sliding windows).\label{fig:cp2}}
\end{figure*}

\begin{figure}[h]
\centerline{\includegraphics[width=\columnwidth]{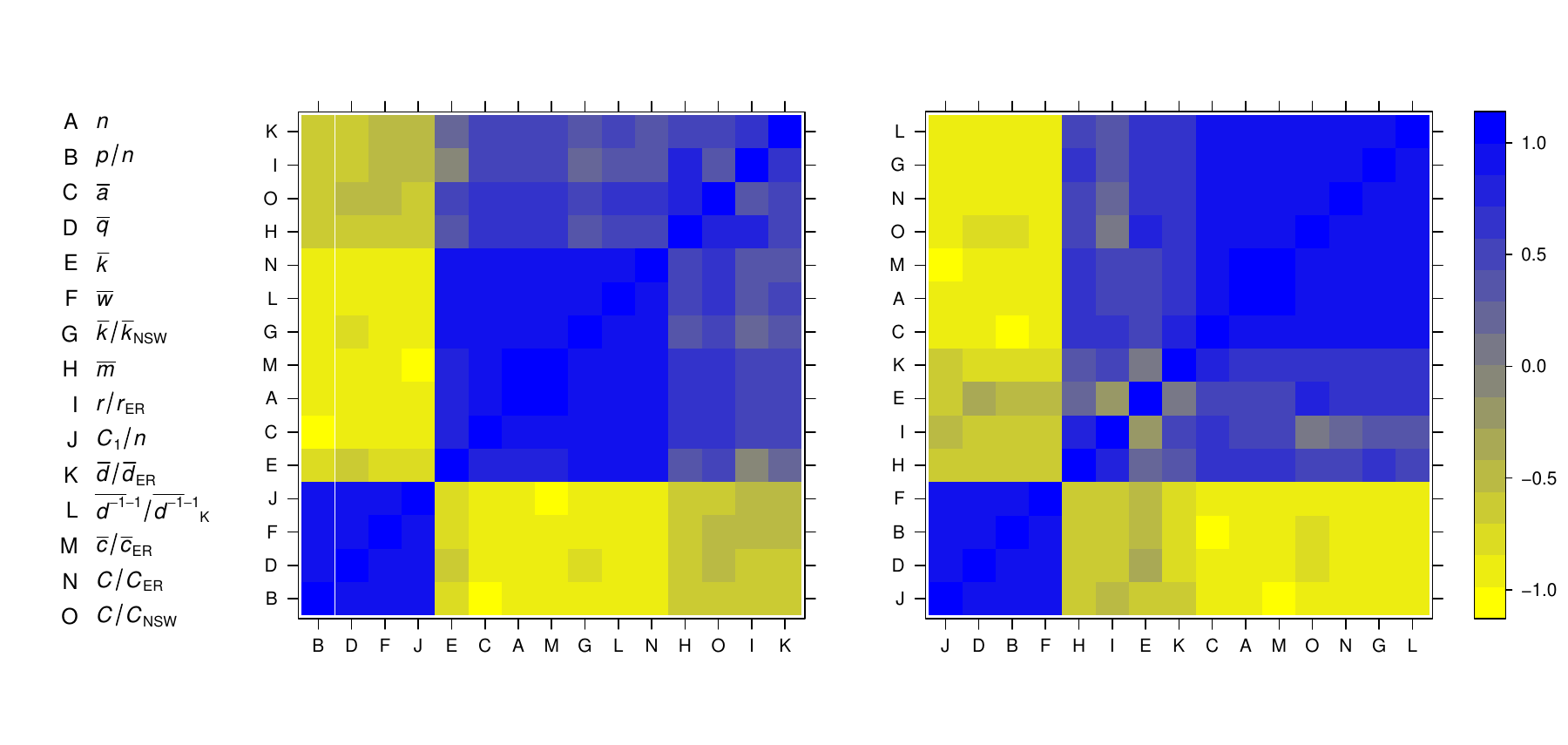}}
\caption{\small Correlation matrices of time series of key statistics across 5-year sliding windows (left) on the complete MR network and (right) on the few-author subnetwork.\label{fig:corrplot}}
\end{figure}

We grouped the time series by shade (color online) in Figs.~13 and 14 (main text) and in Figs.~\ref{fig:cp1} and \ref{fig:cp2} (supplementary materials) according to the strength of their correlations.
Fig.~\ref{fig:corrplot} depicts correlation plots for the original time series {\sc(a)} and for those computed for the few-author network {\sc(b)}.

\begin{figure}[h]
\centerline{\includegraphics[width=1\columnwidth]{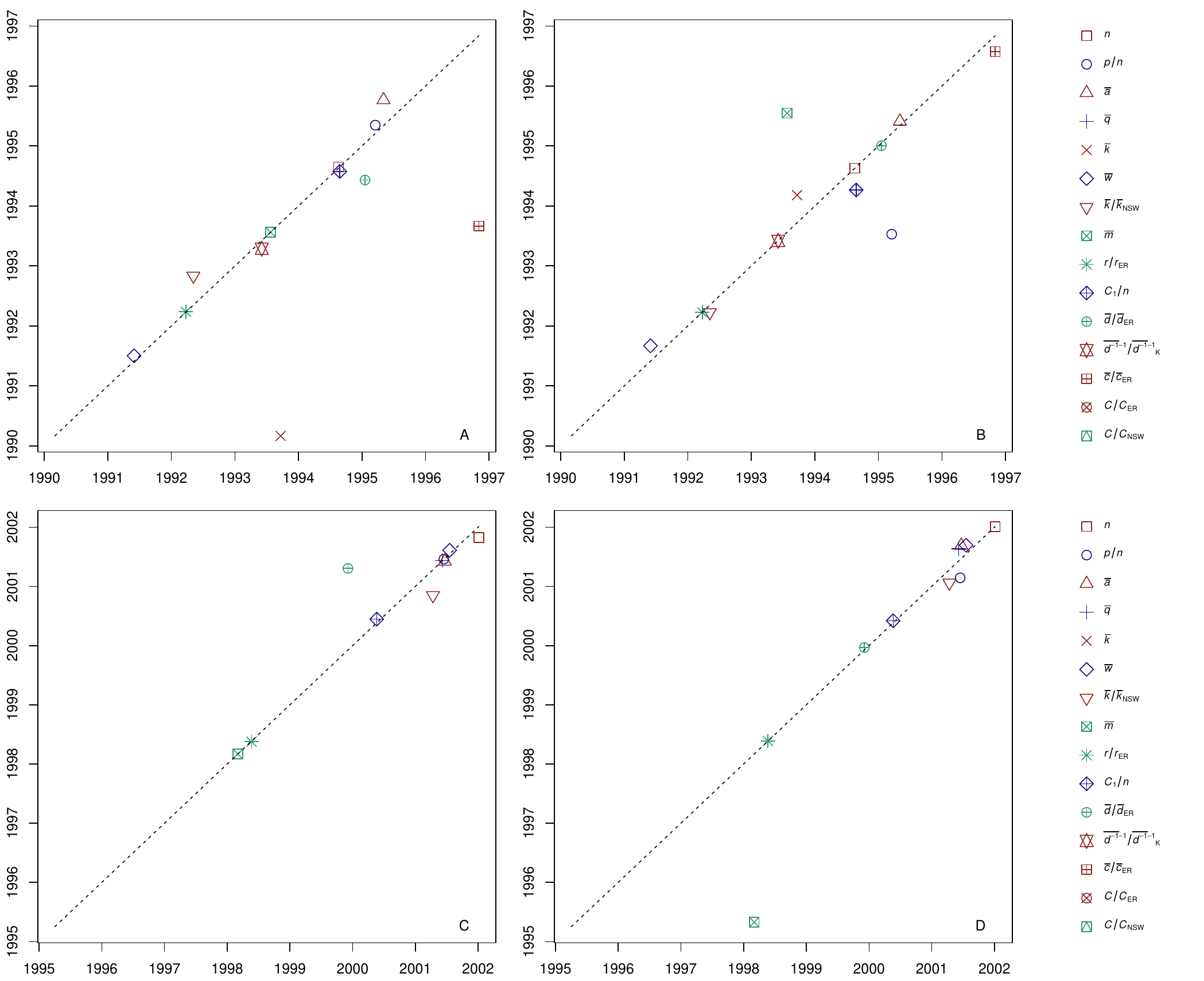}}
\caption{\small Delay plots for both change points using time series across a 3-year sliding window.\label{fig:delaywin3}}
\end{figure}

We performed the same analysis---including time series and their normalizations for all diagnostics---over a 3-year sliding window for comparison.
The time series closely track those across a 5-year window.
We present in Fig.~\ref{fig:delaywin3} the delay plots here for visual reference.

